\documentclass[onecolumn,amsmath,amssymb,superscriptaddress,nofootinbib,11pt,a4paper]{revtex4}

\pdfoutput=1
\usepackage[T1]{fontenc}
\usepackage{CJK}
\usepackage{xcolor}
\usepackage{amsfonts}
\usepackage{epstopdf}
\usepackage{wrapfig} 
\usepackage{subfigure}
\usepackage{graphicx}  
\usepackage{dcolumn}   
\usepackage{bm}
\usepackage{float}
\usepackage[T1]{fontenc}
\usepackage{CJK}
\usepackage{xcolor}
\usepackage{amsfonts}
\usepackage{epstopdf}
\usepackage{wrapfig} 
\usepackage{subfigure}
\usepackage{dcolumn}   
\usepackage{bm}
\usepackage{float}
\usepackage{braket}

\begin{document}
\title{Holographic study on Einstein ring for a charged black hole in conformal gravity}
\author{Xin-Yun Hu}
\affiliation{College of Economic and Management, Chongqing Jiaotong University, Chongqing 400074, China}
\author{Xiao-Xiong Zeng}\email{xxzengphysics@163.com}
\affiliation{Department of Mechanics, Chongqing Jiaotong University, Chongqing 400074, China}
\author{Li-Fang Li}{}\thanks{Corresponding author: lilifang@imech.ac.cn}
\affiliation{Center for Gravitational Wave Experiment, National Microgravity Laboratory, Institute of Mechanics, Chinese Academy of Sciences, Beijing 100190, China.}
\author{Peng Xu}
\affiliation{Center for Gravitational Wave Experiment, National Microgravity Laboratory, Institute of Mechanics, Chinese Academy of Sciences, Beijing 100190, China.}
\affiliation{Lanzhou Center of Theoretical Physics, Lanzhou University, No. 222 South Tianshui Road, Lanzhou 730000, China}

\begin{abstract}
{Under the help of AdS/CFT correspondence, the Einstein ring of a charged black hole in conformal gravity has been studied. Imposing an oscillating Gauss source on one side of the AdS boundary which propagates in the bulk, the response function are derived on the other side of the boundary. Given the proposed wave optics system, the Einstein rings are observed as expected. All these results reveal that when the observer locates at the north pole, there always exists the Einstein ring and surrounding concentric stripes. And such ring becomes into a luminosity-deformed ring or light spot while the observer departs away from the north pole.
We also investigate the effect of temperature $T$, chemical potential $u$ and 
gravity-related parameters $c_0$ on the ring radius. We find the ring radius increases with the decrease of the temperature, increase of the chemical potential, and increase of the gravity-related parameters respectively. 
To check the results in the framework of holography, we investigate the ingong angle of photon at the photon ring via geometric optics and find it is consistent with the angle of Einstein ring obtained via holography.}
\end{abstract}

\maketitle
 \newpage
\section{Introduction}
Over the last century, Einstein's General Relativity (GR) had passed through numerous tests~\cite{Giesler2019,Isi2019,Isi2021,Isi_2021,Franchini2023}, and acquired significant observational supports.
Therefore, GR becomes the theoretical foundations of a wide range of research fields, including cosmology, astrophysics, gravitational wave detection, unified field theory etc. \cite{Bamber2021,Cardoso2022,Barack2019,Meszaros2019_astro}. While, challenges still exist from the observations at larger, galactic scale. The long-standing theoretical issues, like the resolutions of singularities and searches for quantum theory of gravity \cite{Ashtekar2014,Pawlowski2021}, turn out to be much severer and more critical at the beginning of this new century of relativity. 

As we all know, shadows are expected to be found in black holes~\cite{Akiyama2019,Akiyama2019Lett,Akiyama2019Astrophys,Akiyama2019_Lett,Akiyama2019_ring,Akiyama2019_875}. To image this phenomenon, the Event Horizon Telescope, a global long baseline interference array observing at a wavelength of
1.3 mm, is able to reconstruct event-horizon-scale images of the supermassive black hole candidate in the center of the giant elliptical galaxy M87$^\star$. 
And the black hole shadow and its observation along with other significant properties have also been analyzed in~\cite{Gralla2019,Zeng2023,Zeng2020,Zeng:2020vsj}. In these days, a number of researchers have devoted themselves to exploring the physics of black hole images.

In the papers~\cite{
Hashimoto:2018okj,Hashimoto:2019jmw}, they proposed a direct procedure to construct holographic images of the black hole in the
bulk from a given response function of the QFT on the boundary with the AdS/CFT duality, which establishes a connection between gravity and field theories. As we know, the realization of the holographic principle is motivated by string theory. This duality in theoretical physics connects two seemingly different theories: the gravitational theory in a (d+1)-dimensional Anti-de Sitter space and a d-dimensional conformal field theory. With this AdS/CFT correspondence, the response function with respect to an external source corresponds to the asymptotic data of
the bulk field generated by the source on the AdS boundary. Using this response function, the holographic images gravitationally lensed by the black hole can be constructed for a thermal state on two-dimensional sphere which is dual to Schwarzschild AdS$_4$ black hole. The obtained results are consistent with the size of the photon sphere of the black hole calculated through the geometrical optics method. Following these breakthroughs~\cite{Liu:2022cev,Zeng:2023zlf,
Zeng_Li2023,CEJM,Hu:2023eow}, the results showed that these holographic images exist in different gravitational background. But according to the specific bulk dual geometry, the photon sphere varies and  the detailed behavior of Einstein ring varies. Therefore, in this paper, we are tempted to investigate the behavior of the lensed response for the charged black hole in conformal gravity. On one hand, we want to investigate the effect of the chemical potential on the Einstein ring. As stressed in \cite{Liu:2022cev}, the   real quantum materials are engineered  at a
finite chemical potential, the chemical potential plays an important role in strongly coupled systems. Therefore, investigation on the effect of the chemical potential on the Einstein ring is also important and necessary. On the other hand, we intend to explore the effect of the gravity-related parameter in conformal gravity on the Einstein ring.  And whether we can distinguish the conformal  gravity from Einstein  gravity via holographic images of black holes. 
The action of 
conformal gravity is defined
 by the Weyl tensor square, which allows for more solutions
than Einstein gravity. Especially, it was  regarded as a desirable UV
culmination of gravity \cite{1} and useful setting up of supergravity theories \cite{2}. Moreover in conformal gravity, dark matter and dark energy are not necessary in explaining the galaxy rotation curves \cite{3,4}. 
Therefore, conformal gravity is  an important alternative gravity and thus 
investigations on the Einsteins ring in this gravity are also important and interesting. 

This paper is organized as follows. In section~\ref{sec2}, we review the charged black hole solution in conformal gravity for a spherically symmetric AdS black hole. In section ~\ref{sec3}, we give a holographic setup of such model and analyze the corresponding lensed response function, explicitly. Given the optical system, we indeed observe the Einstein ring in our model and compare our results with the optical approximation in section~\ref{sec4}. Our results shows that the position of photon ring obtained from the geometrical optics is full consistence with that of the holographic ring. Section~\ref{sec5} is our conclusions.

\section{Review of the holographic construction of Einstein ring in AdS black holes}
\label{sec2}

Conformal pure gravity is constructed from the Weyl-squared term, and its conformal symmetry is preserved when it couples minimally to the Maxwell field. Its Lagrangian is ~\cite{Li:2012gh}
\begin{equation}
h^{-1} \mathcal{L}=-\frac{1}{4}C^{\mu\nu\rho\sigma}C_{\mu\nu\rho\sigma}-\frac{1}{6}F^2,
\end{equation}
here $h=\sqrt{g}$ and $g$ is the determinant of the metric. $F=dA$ and $C_{\mu\nu\rho\sigma}$ is the Weyl tensor. The corresponding equations of motion are
\begin{equation}
\nabla^{\mu}F_{\mu\nu}=
\frac{1}{2}(2\nabla^{\rho}\nabla^{\sigma}+R^{\rho\sigma})C_{\mu\rho\sigma\nu}+\frac{1}{3}(F^2_{\mu\nu}-\frac{1}{4}F^2g_{\mu\nu})=0.
\end{equation}
  The static charged AdS black hole solution in conformal gravity is~\cite{Li:2012gh}
\begin{equation}
ds^2=\text{} -F(r)dt^2 + \text{}\frac{1}{F(r)} dr^2\text{} + r^2 d\Omega_{2,\epsilon}^2,
\label{metric_metric}
\end{equation}
with
\begin{align}
F(r)= c_0+c_1 r -\frac{1}{3} \Lambda r^2+\frac{d}{r} \label{fr_fr},
\end{align}
and the Maxwell field
\begin{align}\label{eq4}
A=-\frac{Q}{r}\text{} dt.
\end{align}

The parameter $\epsilon$ can take three discrete values, -1, 0, 1, which correspond to the hyperbolic, planar and spherical geometry respectively. The solutions with $\epsilon=-1,0$ are the topological black holes, which exist only in AdS backgrounds. In this paper, we are interested in the case $\epsilon=1$. The integration constants 
$Q,c_0, c_1, d, \Lambda$ obey the constraint condition 
\begin{align} 
3 c_1  d+\epsilon^2+Q^2-c_0^2=\text{}0,
\label{constrain}
\end{align}
in which $c_1$ is the massive spin -2 hair,  $Q$  is regarded as the electric charge and  $d$ is regarded as the mass. Note that in Eq.(\ref{fr_fr}), the charge $Q$ does not appear as the usual charged black hole. It  affects
the geometry of the spacetime only implicitly through the constraint condition shown in Eq.(\ref{constrain}). 
In addition $Q$
appears only in squared form in Eq.(\ref{constrain}), the spacetime geometry does not
discriminate positive and negative values of $Q$. In this paper, we will consider
exclusively the case $Q\geq0$ as in~\cite{Li:2012gh}.
For the case $ \Lambda <0$, there is only a nonzero real positive root of equation $F(r)=0$, which corresponds to the event horizon of the back hole. 
The solution describes a charged AdS black hole in this case, and we set $1/l^2= -\frac{1}{3} \Lambda $, where $l$  is the Ads radius. 

To reduce the parameters, we will keep  $c_0$ while drop  $c_1$.  From Eq.(\ref{constrain}), we can get 
\begin{align}
c_1=\frac{ c_0^2-Q^2-1}{3 \text{}d}\label{constrain1}.
\end{align}
Substituting this condition into Eq.(\ref{fr_fr}), we have 
\begin{align}
F(r)= c_0+\frac{ c_0^2-Q^2-1}{3 d} \text{}r -\frac{1}{3} \Lambda r^2+\frac{d}{r}, 
\label{Fr_3}
\end{align}
where we have set $l=1$. The Hawking temperature in this case is 
\begin{align}
T=\frac{\kappa}{2 \pi}=\frac{{F(r)}^{\prime}}{4 \pi}\mid_{r_h}=\frac{6 r_h^3-\sqrt{3} \sqrt{r_h^2 \left(-c_0^2+6 c_0 r_h^2+3 r_h^4+4 Q^2+4\right)}}{12 \pi  \text{}r_h^2}\label{tm},
\end{align}
in which $\kappa$ is the surface gravity and  $r_h$ is the event horizon of the black hole. 

Next we focus on the spherical metric (\ref{Fr_3}). With a new definition $r_*=1/r$, we have $F(r)=r_*^{-2}F(r_*)$. We rewrite the metric (\ref{metric_metric}) in the new coordinate $(t, r_*, \theta, \phi)$ and derive
\begin{equation}
ds^2=\frac{1}{r_*^2}[-F(r_*)dt^2+\text{}\frac{d r_*^2}{F(r_*)}+d\Omega^2],
\label{metric_2}
\end{equation}
here $d\Omega^2=d\theta^2+\sin 
\theta^2 d\phi^2$. We will take Eq.(\ref{metric_2}) as our background.

The complex scalar field $\Phi$ in such background is determined by the following Klein-Gordan equation~\cite{Liu:2022cev}
\begin{equation}
D_b D^b \Phi \text{}- M^2\Phi=0,
\end{equation}
here $D_a=\nabla_a -i e A_a$ and we will take $M^2=-2$ in the following. In order to solve the above Klein-Gordon equation easily, the ingoing Eddington coordinate is introduced, which is expressed as
\begin{equation}
\nu=t-\int \frac{\text{}d r_*}{F(r_*)}\text{}.
\end{equation}
With the above new coordinate, the metric (\ref{metric_2}) is further rewritten as
\begin{equation}
ds^2=\frac{1\text{}}{r_*^2}[-F(r_*)d\nu^2\text{}-2 d r_* d\nu+d\Omega^2].
\end{equation}
The asymptotic solution of the complex scalar field $\Phi$ near the AdS boundary becomes \cite{Liu:2022cev}
\begin{equation}\text{}\text{}\text{}\text{}
\Phi(\nu,r_*,\theta,\phi)=r_* J_{\mathcal{O}}(\nu,\theta,\phi)+r_*^2 \text{}\langle \mathcal O \rangle+\mathrm{O}(r_*^3).\text{}\text{}\text{}\text{} \text{}
\end{equation}
According to the AdS/CFT correspondence, $J_{\mathcal{O\text{}}}$ is the source for the boundary field theory. And the corresponding expectation value of dual operator is 
\begin{equation}
\langle\text{} \mathcal O \rangle_{J_{\mathcal{O}}}=\langle \mathcal O \rangle-(\partial_{\nu}\text{}-ie u) J_{{\mathcal{O}}},
\end{equation}
here $\langle \mathcal{O} \rangle_{J_{\mathcal{O}}}$ is also named the response function, in which $u= Q/r_h$ is the chemical potential. Obviously, $\langle\mathcal{O}\rangle$ corresponds to the expectation value of the dual operator with the source turned off.  

\begin{figure}[ht]
	\centering
\includegraphics[trim=3.2cm 2.2cm 1.9cm 0.1cm, clip=true, scale=0.9]{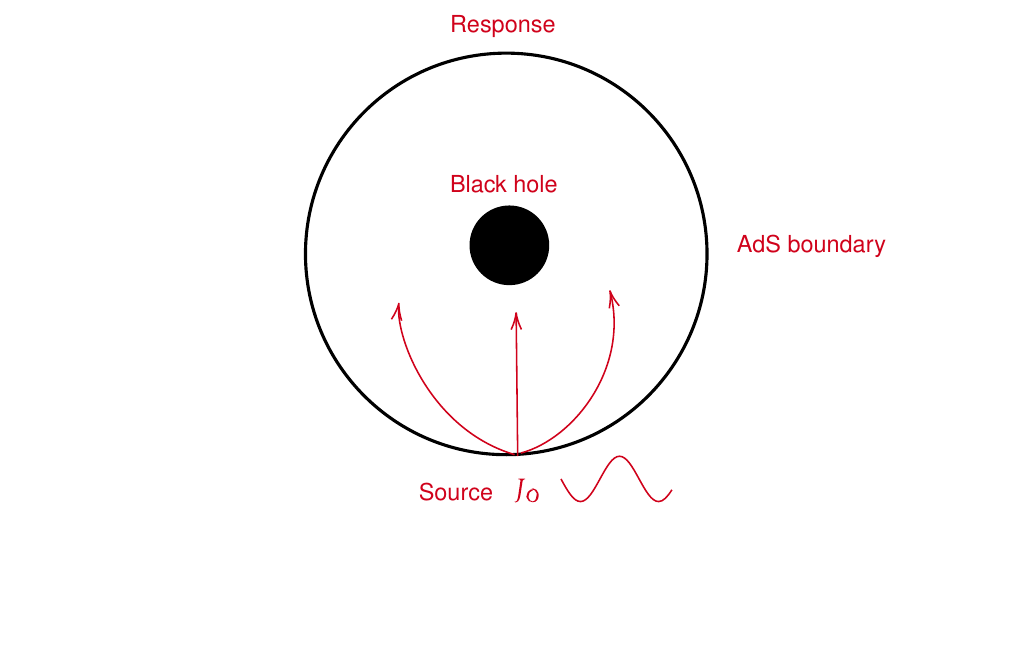}
\caption{A monochromatic Gaussian source is at one side of the AdS boundary, and its response is observed at another side on the same boundary. }\label{source_and_response}
\end{figure}

Here we would like to study the holographic Einstein images of the charged black hole in conformal gravity based on the AdS/CFT correspondence. We choose an oscillatory Gaussian wave source $J_{\mathcal{O}\text{}}(\text{}\nu,\theta)$ on one side of the AdS boundary, and the scalar waves generated by the source propagate in the bulk which is shown in Fig. ~\ref{source_and_response}. When the bulk of scalar waves reach the other side of the AdS boundary, the response function $\langle\text{} \mathcal {O} \rangle_{J_{{\mathcal{O}}}\text{}}$ is generated on the opposite side of the source. Here we take a monochromatic and axisymmetric Gaussian wave packet which is centered on the south pole $\theta_0\text{}=\pi\text{}$ as the source
\begin{eqnarray}
\text{}J_{\mathcal{O}}(\nu,\text{}\theta)&=&\frac{1}{2\pi\sigma^2 }e^{-i \omega\text{} \nu}\exp{[-\frac{(\pi-\theta)^2}{2\sigma^2}]}\nonumber \\
&=&e^{-i \omega \nu} \sum_{l=0}^{\infty}C_{l0}Y_{l0}(\theta),
\end{eqnarray}
$\sigma$ is the wave width which is produced by Gaussian source. $Y_{l0}$ is the spherical harmonics function. If we take the size $\sigma<<\pi$, the corresponding coefficients of $Y_{l0}$ is expressed as
\begin{eqnarray}
C_{l0}=(-1)^l (\frac{l+1/2}{2\pi})^{1/2}\exp{[-\frac{(l+1/2)^2\tau^2}{2}]}.
\end{eqnarray}

Considering the spacetime symmetry, the complex scalar field $\Phi(\nu,r_*,\theta,\phi)$ is further decomposed as
\begin{eqnarray}
\Phi(\nu,r_*,\theta,\phi)=e^{-i\omega \nu}\sum_{l=0}^{\infty}C_{l0}r_{*}\mathcal{Z}_l(r_*)Y_{l0}(\theta),
\end{eqnarray}
where $\mathcal{Z}_l$ satisfies the equation of motion
\begin{equation}
r_*^2 f \mathcal{Z}_l''+r_*^2[f^{\prime}+2 i (\omega-e A)]\mathcal{Z}_l^{\prime}+[(2-\text{}2f)+ r_*f^{\prime}-r_*^2\text{} l(l+1)-i e r_*^2 A^{\prime}]\mathcal{Z}_l=0,
\end{equation}
where $e$ is the charge of the complex scalar field. The asymptotic behavior of $\text{}\mathcal{Z}_l$ near the AdS boundary goes like
\begin{equation}
\text{}\mathcal{Z}_l=1+ r_* \langle \mathcal{O}_l\text{}\rangle+O(r_*^2).
\end{equation}
And the response function $\langle\mathcal{O}\rangle_{\text{}J_{\mathcal{O}}}$ is also expressed as
\begin{equation}
\langle\mathcal{O\text{}}\rangle_{J_{\mathcal{O}}}=e^{-i\omega \text{}\nu}\sum_{l=0}^{\infty} C_{l0} \langle \mathcal{O}\text{}\rangle_{J_{\mathcal{O}l}}Y_{l 0}(\theta).
\label{total_response}
\end{equation}
Then we have
\begin{equation}
\langle \mathcal{O }\rangle_{J_{\mathcal{O}l}}=\langle \mathcal{O}\rangle_l+i\hat {\omega},
\end{equation}
here $\hat {\omega}=\text{}\omega+e u$.

\begin{figure}
    \centering
    \includegraphics[width=4in]{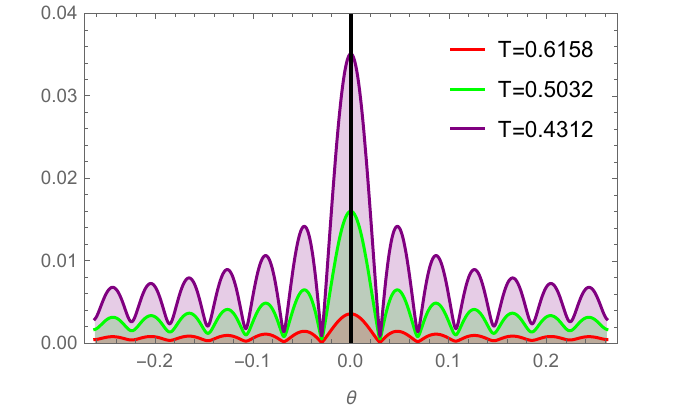}
    \caption{The amplitude of $\langle\mathcal{O}\rangle$ with different temperature $T$ for $Q=0.5$, $\omega=80$, $c_0=0.5$ and $e=1$.}
    \label{profile1}
\end{figure}

\begin{figure}
    \centering
    \subfigure{
        \includegraphics[width=4in]{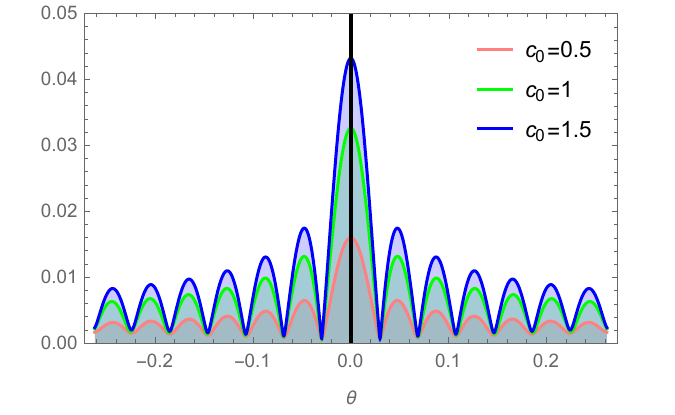}
    }
    \caption{The amplitude of $\langle\mathcal{O} \rangle$ with different $c_0$ for $r_{*h}=0.5$, $Q=0.5$, $\omega =80$ and $e=1$.}
    \label{profile2}
\end{figure}

\begin{figure}
    \centering
    \subfigure{
        \includegraphics[width=4in]{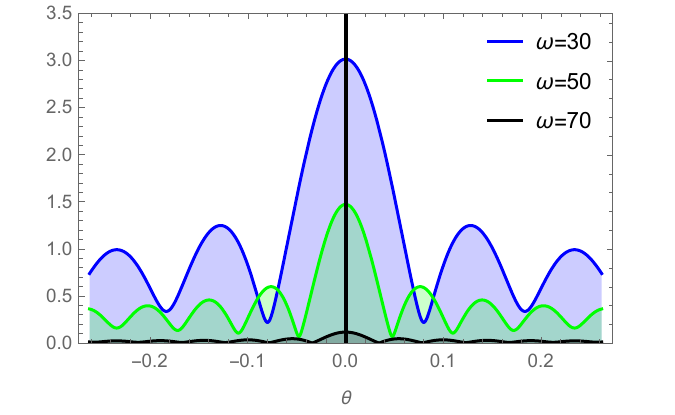}
    }
    \caption{The amplitude of $\langle\mathcal{O}\rangle$ with different $\omega$ for $r_{*h}=0.5$, $c_o=0.5$, $Q =0.5$ and $e=1$.}
    \label{profile3}
\end{figure}

\begin{figure}
    \centering
    \subfigure{
        \includegraphics[width=4in]{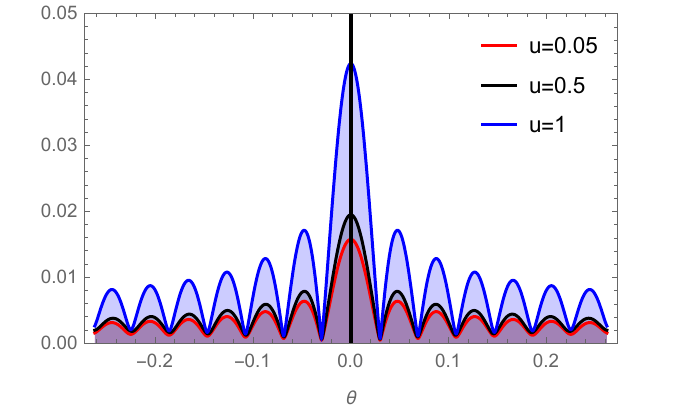}
    }
    \caption{The amplitude of $\langle\mathcal{O}\rangle$ with different $u$ for $r_{*h}=0.5$, $\omega=80$, $c_0=0.5$ and $e=1$.}
    \label{profile4}
\end{figure}

\begin{figure}
    \centering
    \subfigure{
        \includegraphics[width=4in]{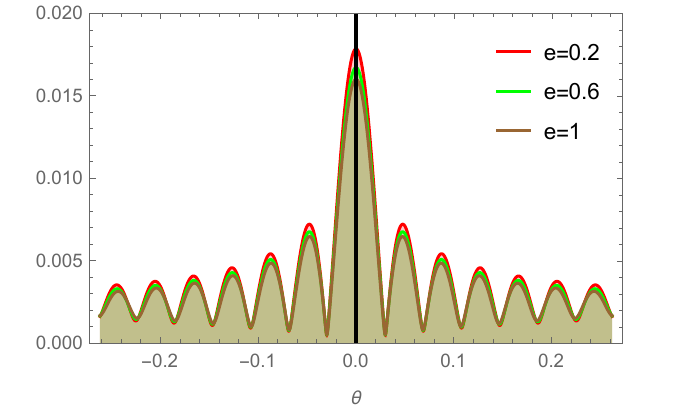}
    }
    \caption{The amplitude of $\langle \mathcal{O}\rangle$ with different $e$ for $Q=0.5$, $\omega=80$, $c_0=0.5$, and $r_{*h}=0.5$.}
    \label{profile5}
\end{figure}

Under the help of the pseudo-spectral method, we derive the numerical solution~\cite{Hashimoto:2018okj,Hashimoto:2019jmw} for $\mathcal{Z}_l$ and $\mathcal{O}\text{}_l$ straightforwardly. With  $\mathcal{O\text{}}_l$, we obtain the total response by Eq.(\ref{total_response}). The typical profiles of the total response are shown in Fig.~\ref{profile1} to Fig.~\ref{profile5}. We indeed see the interference pattern produced from the diffraction of our scalar field by the black hole. Explicitly in Fig.~\ref{profile1}, we plot the amplitude of $\langle O\text{} \rangle$ for different $T$  with $Q=0.5$, $\omega=80$, $c_0=0.5$ and $e=1$. We clearly see that when the temperature increases, while the amplitude of the total response function decreases. Also we plot the amplitude of $\text{}\langle O \rangle$ for different $c_0$ with $r_{*h}=0.5$, $Q=0.5$, $\omega=80$ and $e=1$ in Fig.~\ref{profile2}. The amplitude of the total response function raises  with the increasing of the values 
 of parameter $c_0$. Then we plot the amplitude of $\langle O \rangle$ for different $\omega$ with $r_{*h}=0.5$, $c_0=0.5$, $Q=0.5$ and $e=1$ in Fig.~\ref{profile3}. The oscillation period of the wave is maximum when $\omega=30$ and the amplitude decreases as $\omega$ increases to $\omega=50$ and $\omega=70$. The results mean the frequency $\omega$ of the Gaussian source can increase the wave width and reduce the wave period. These in turn shows the total response function depends closely on the Gaussian source. We further plot the amplitude of $\langle O \rangle$ for different chemical potential $u$ with $r_{*h}=0.5$, $\text{}\omega=80$, $\text{}c_0=0.5$ and $e=1$ in Fig.~\ref{profile4}. When the chemical potential is small, $u=0.05\text{}$ ,  the amplitude of $\langle O \rangle$ is also small. When the chemical potential gradually increases to $u=1$, the amplitude also gradually increases. At last, we show the amplitude of $\langle O \rangle$ for different charge $e$ with   parameters $Q=0.5$, $\text{}\omega=80$, $c_0=0.5$ and $r_{*h}=0.5$ in Fig.~\ref{profile5}. We see that the change of the amplitude  $\langle O \rangle$ in this case seems not obvious as in other cases, although we can observe that the larger the value of  $e$, the smaller the  value of  amplitude $\langle O \rangle$.
 
\section{The holographic ring formation setup}
\label{sec3}
\begin{figure}[ht]
	\centering
	\includegraphics[height=3.0in]{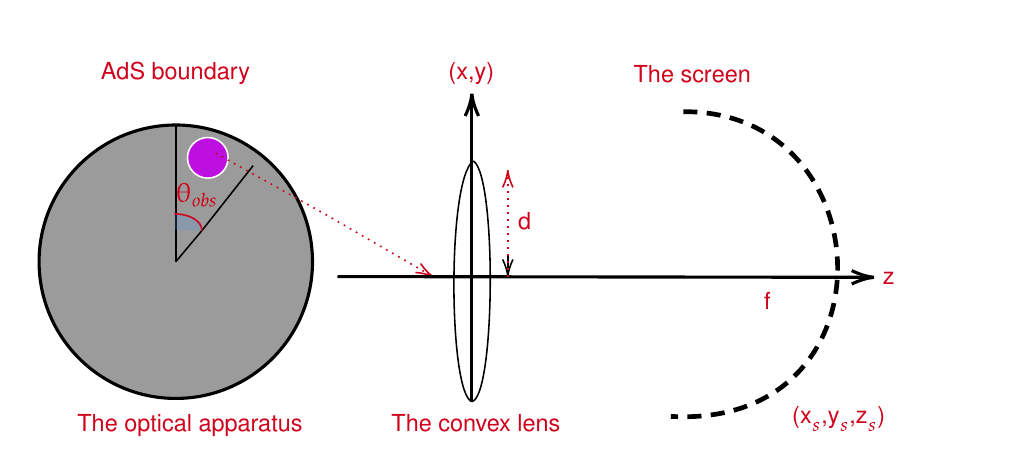}
	\caption{The optical system with a convex lens and a telescope. }\label{optics_new}
\end{figure}

As stated in~\cite{Liu:2022cev}, we need an optical system to observe the response function which is shown in Fig.~\ref{optics_new}. An optical apparatus, a convex lens and a spherical screen are needed.  Here we suppose the lens is infinitely thin. The size of the lens is much smaller than the focal length $f$. And the lens is regarded as a “converter” between
plane and spherical waves. The right side of the optical system is a screen and the received image from the transmitted wave is depicted on this screen.

Here we consider a wave $\Psi(\hat{x})$ irradiated to the lens from the left side. Such wave is converted to $\Psi_T(\hat{x})$ which converges at the focus $f$. The relationship between $\Psi(\hat{x})$ and $\Psi_T(\hat{x})$ is 
\begin{equation}
\Psi_T(\hat{x})=e^{-\frac{i {\omega} |\hat{x}|^2}{2f}}\Psi(\hat{x}).
\end{equation}
Fixing a spherical screen at a cartesian coordinate system $(x,y,z)=(x_s,y_s,z_s)$ with $x_s^2+y_s^2+z_s^2=f^2$~\cite{Hashimoto:2018okj,Hashimoto:2019jmw}, the transmitted wave function $\Psi_T(\hat{x})$ imaging onto the screen is changed into a new function $\Psi_s(\hat{x})$, which can be expressed as 
\begin{eqnarray}
\Psi_s(\hat{x})&=&\int_{|\hat{x}|\leq d} \Psi_T(\hat{x}) e^{i {\omega} L} d^2 x\nonumber\\
&=&\int_{|\hat{x}|\leq d} \Psi(\hat{x}) e^{-\frac{i {\omega}\hat{x}\cdot\hat{x_s}}{f}} d^2 x\nonumber\\
&=&\int \mathrm{w}(\hat{x}) \Psi(\hat{x}) e^{-\frac{i {\omega}\hat{x}\cdot\hat{x_s}}{f}} d^2 x,
\label{q_1}
\end{eqnarray}
where $d$ is the length radius. $L$ is the distance between the lens point $(x,y,0)$ and the screen point $(x_s,y_s,z_s)$. The above window function $\mathrm{w}(\hat{x})$ is given by
\begin{eqnarray}
\mathrm{w}(\hat{x})=\left\{
\begin{aligned}
0&,& 0\leq |\hat{x}|\leq d\\
1&,&|\hat{x}|\geq d
\end{aligned}
\right.
\end{eqnarray}

\begin{figure}[ht]
	\centering
	\includegraphics[trim=1.2cm 0.5cm 4.9cm 2.5cm, clip=true, scale=0.9]{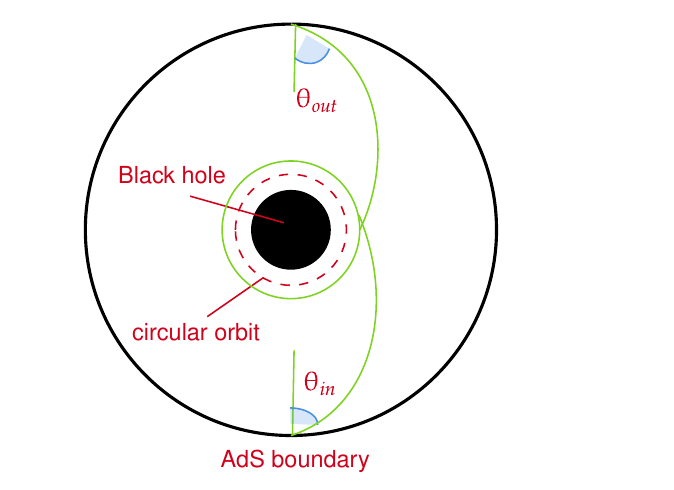}
	\caption{The diagram of the ingoing angle and outgoing angle for a   photon  moving around the black hole.}
 \label{black_source}
\end{figure}

From a wave-optical method, we also have a formula which help us convert the response function $\langle O(\hat{x})\rangle$ to the image of the dual black hole $|\Psi_s(\hat{x}_s)|^2$ on a virtual screen shown as follows
\begin{eqnarray}
\Psi_s(\hat{x}_s)&=&\int_{|\hat{x}|\leq d} \langle O(\hat{x})\rangle e^{-\frac{i {\omega}\hat{x}\cdot\hat{x}_s}{f}}  d^2 x \nonumber\\
&=&\int \mathrm{w}(\hat{x})   \langle O(\hat{x})\rangle e^{-\frac{i {\omega}\hat{x}\cdot\hat{x}_s}{f}}  d^2 x 
\label{q_2}.
\end{eqnarray}

 Next, we will employ Eq.(\ref{q_2}) to construct the Einstein ring and investigate how the wave source and spacetime affect the ring.

\section{Holographic Einstein ring in ADS black hole}
\label{sec4}
\begin{figure}
    \centering
    \subfigure[$\omega=80$]{
        \includegraphics[width=1.4in]{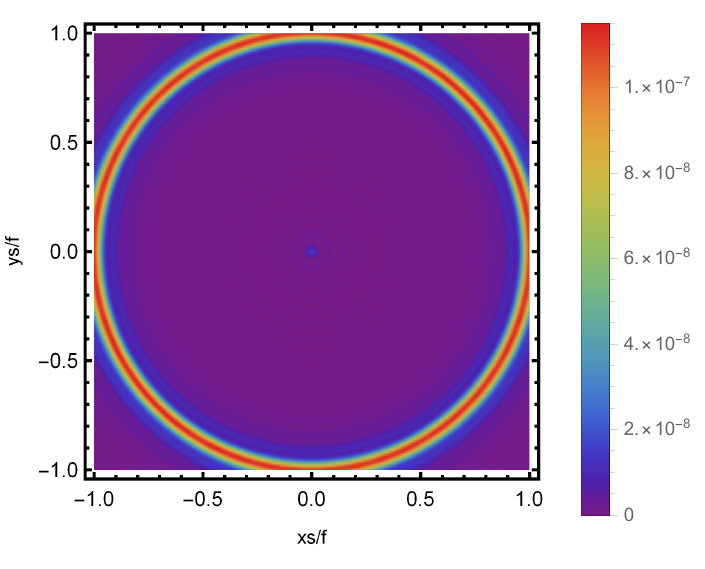}
    }
	\subfigure[$\omega=60$]{
        \includegraphics[width=1.4in]{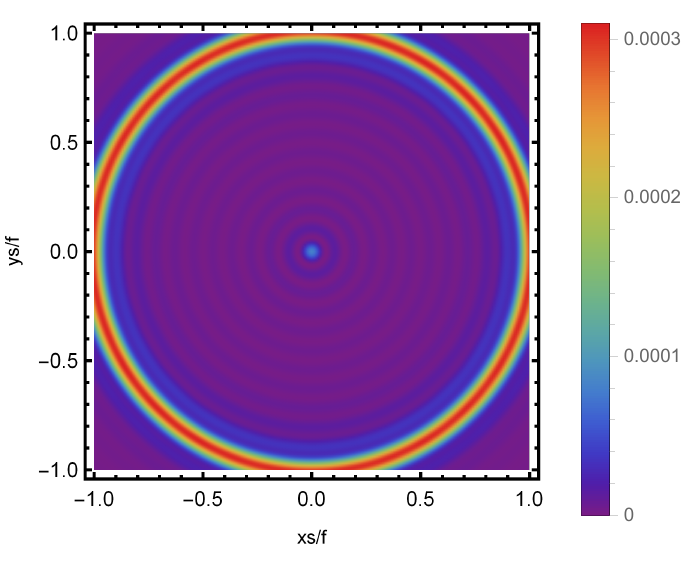}
    }
    \subfigure[$\omega=40$]{
        \includegraphics[width=1.4in]{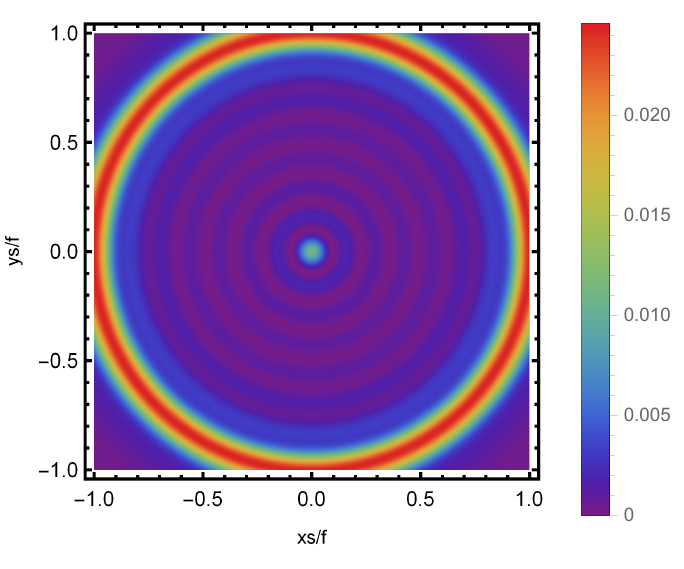}
    }
    \subfigure[$\omega=20$]{
        \includegraphics[width=1.4in]{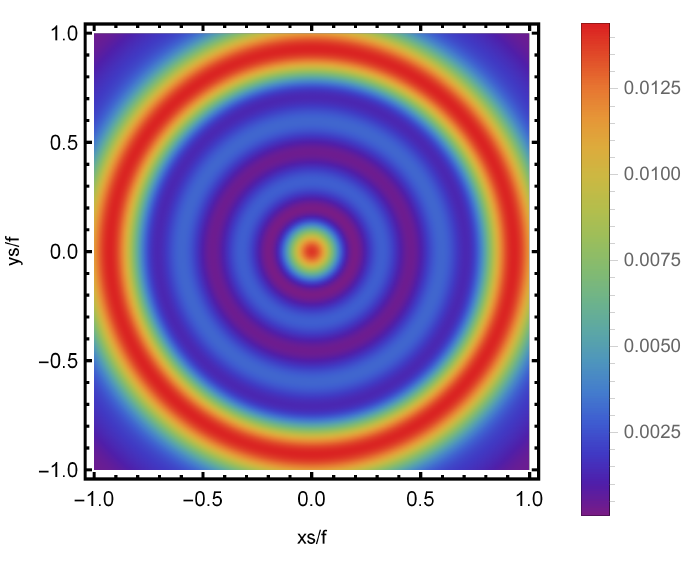}
    }
    \caption{The images of the lensed response observed for different $\omega$ are shown at the observation angle $\theta_{obs}=0$. We see $Q = 0.5$, $r_{*h} = 0.5$, $c_0 = 0.5$, $e = 0.01$.}
    \label{omega_1}
\end{figure}

\begin{figure}
    \centering
	\subfigure[$\omega=80$]{
        \includegraphics[width=1.4in]{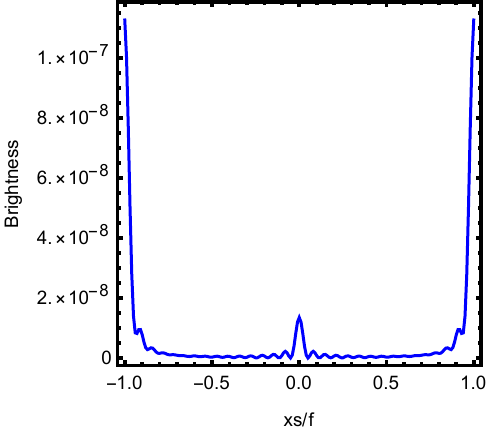}
    }
    \subfigure[$\omega=60$]{
        \includegraphics[width=1.4in]{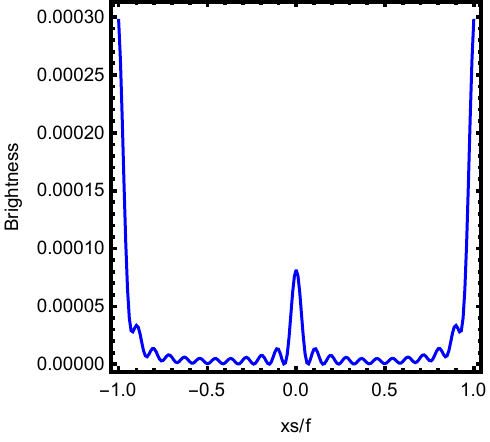}
    }
    \subfigure[$\omega=40$]{
        \includegraphics[width=1.4in]{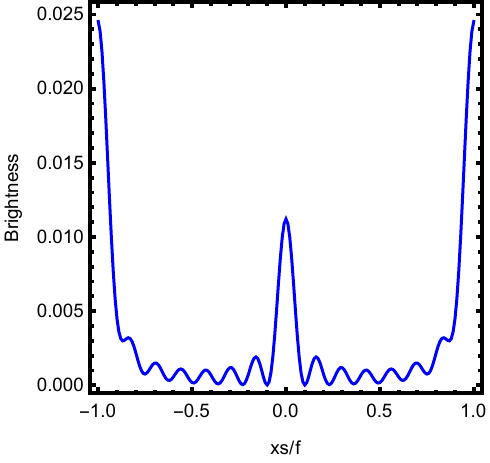}
    }
    \subfigure[$\omega=20$]{
        \includegraphics[width=1.4in]{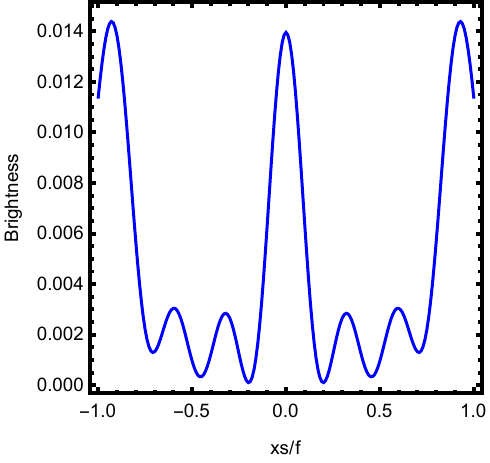}
    }
    \caption{The brightness of the lensed response are shown on the screen for different $\omega$ with $Q = 0.5$, $z_h = 0.5$,  $c_0 = 0.5$, $e = 0.01$.}
     \label{omega_2}
\end{figure}

As predicted on the AdS boundary, we observe different image profiles. In order to study the impact of wave source, we give the holographic Einstein images for various $\omega$ with fixed $Q=0.5$, $r_{*h}=0.5$, $c_0=0.5$ and $e=0.01$ in Fig.~\ref{omega_1} when the position of the observer is at the north pole of the AdS boundary, which means $\theta_{obs}=0$. We clearly see a series of axis-symmetric concentric circular rings appear. And as we increase the frequency $\omega$, the corresponding ring becomes sharper. This means that the geometric optics approximation better describes the image in the high-frequency limit. For a comprehensive understanding of Fig.~\ref{omega_1}, we also plot the profiles of the lensed response function in Fig.~\ref{omega_2}. It is clear that different frequencies lead to changes in the brightness and also the location of brightness peaks.

\begin{figure}
    \centering
    \subfigure[$c_0=1$,$\theta_{obs}=0$]{
        \includegraphics[width=1.4in]{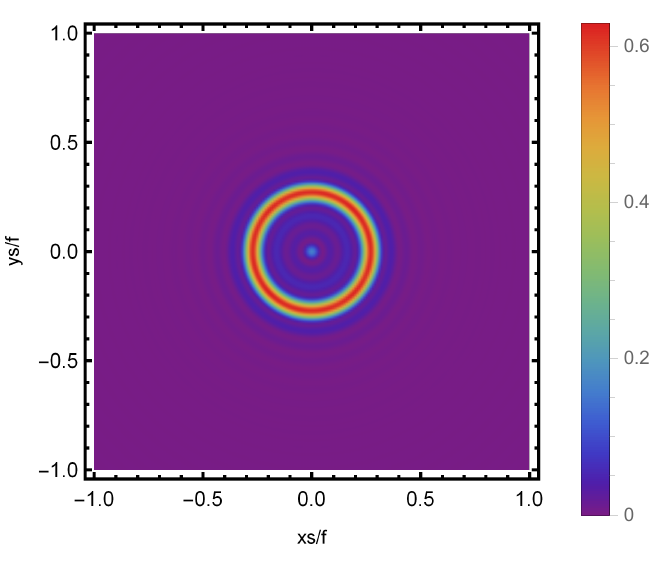}
    }
	\subfigure[$c_0=1$,$\theta_{obs}=\pi/6$]{
        \includegraphics[width=1.4in]{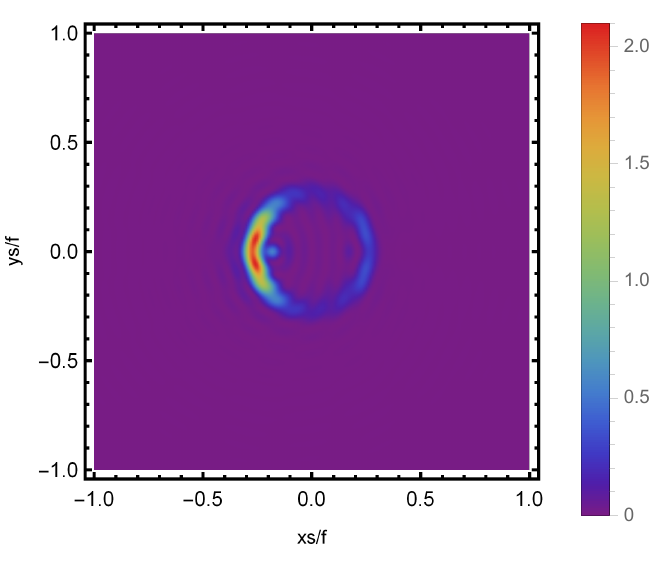}
    }
    \subfigure[$c_0=1$,$\theta_{obs}=\pi/3$]{
        \includegraphics[width=1.4in]{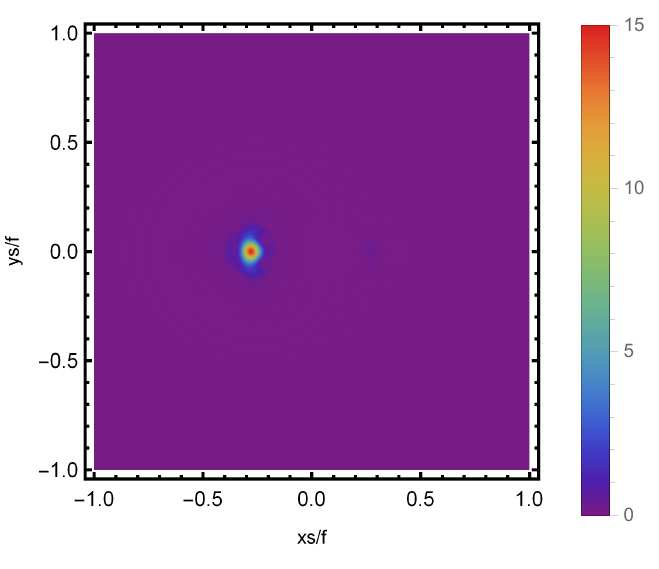}
    }
    \subfigure[$c_0=1$,$\theta_{obs}=\pi/2$]{
        \includegraphics[width=1.4in]{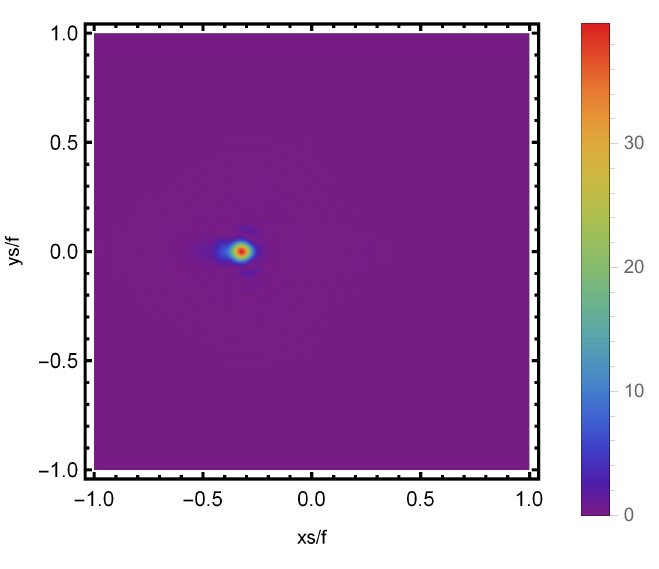}
    }
    \subfigure[$c_0=3$,$\theta_{obs}=0$]{
        \includegraphics[width=1.4in]{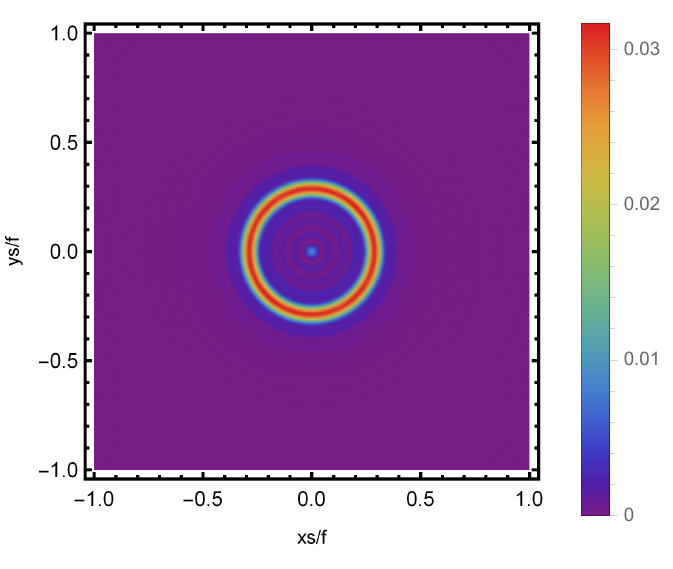}
    }
	\subfigure[$c_0=3$,$\theta_{obs}=\pi/6$]{
        \includegraphics[width=1.4in]{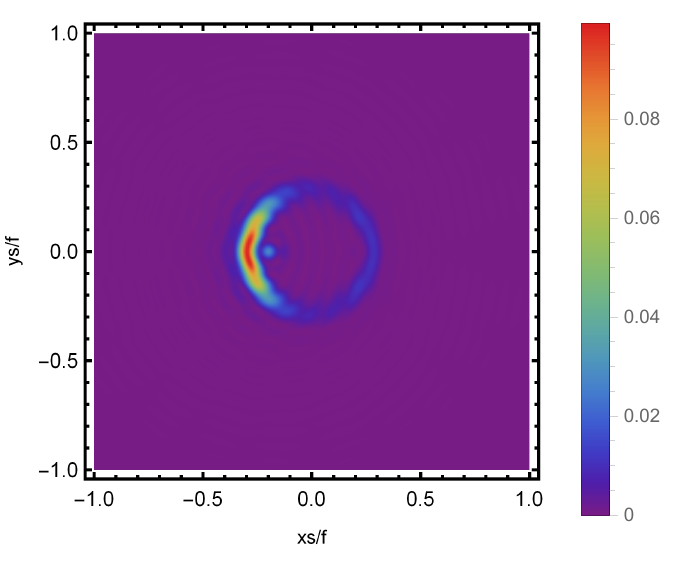}
    }
    \subfigure[$c_0=3$,$\theta_{obs}=\pi/3$]{
        \includegraphics[width=1.4in]{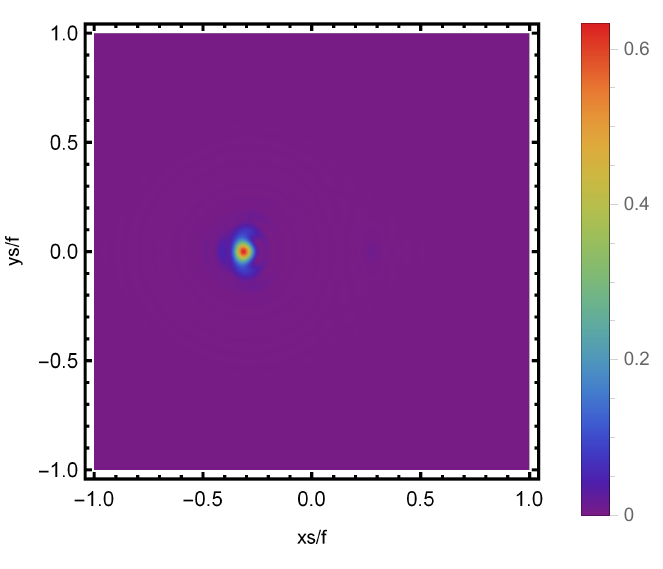}
    }
    \subfigure[$c_0=3$,$\theta_{obs}=\pi/2$]{
        \includegraphics[width=1.4in]{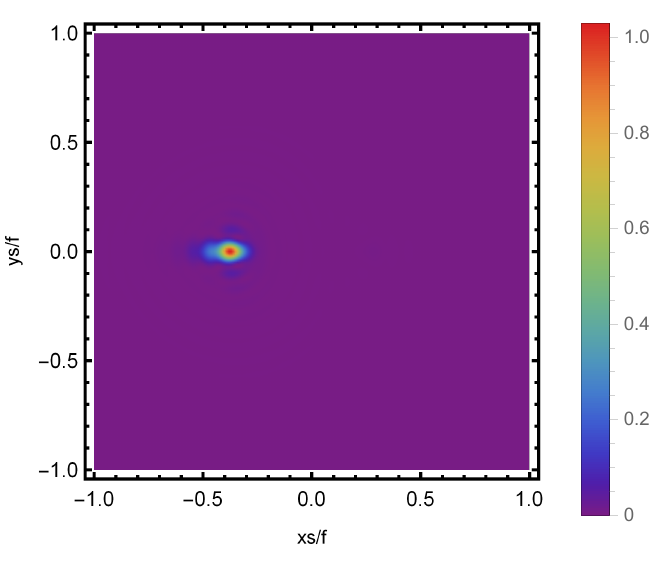}
    }
    \subfigure[$c_0=5$,$\theta_{obs}=0$]{
        \includegraphics[width=1.4in]{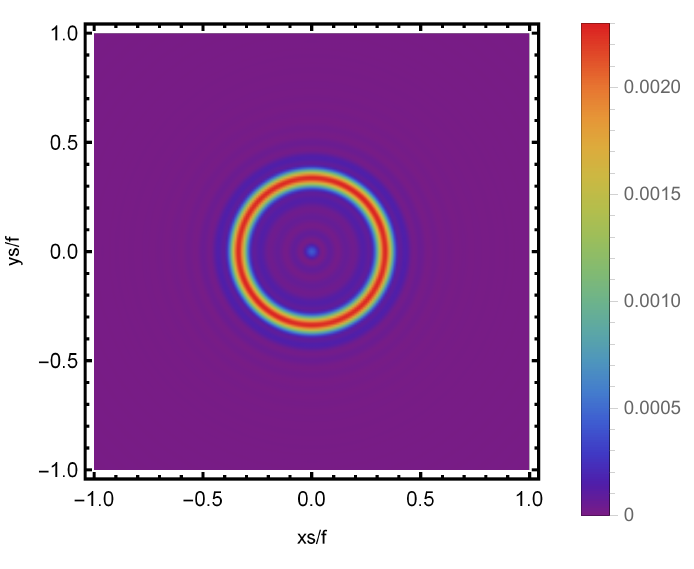}
    }
	\subfigure[$c_0=5$,$\theta_{obs}=\pi/6$]{
        \includegraphics[width=1.4in]{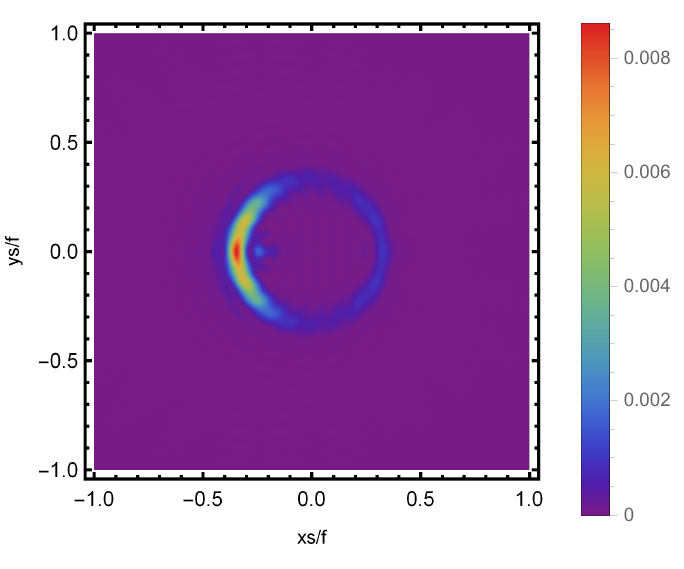}
    }
    \subfigure[$c_0=5$,$\theta_{obs}=\pi/3$]{
        \includegraphics[width=1.4in]{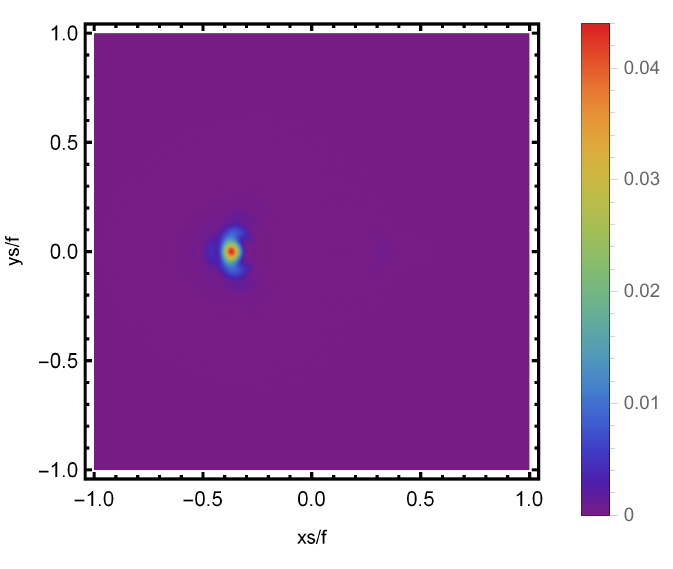}
    }
    \subfigure[$c_0=5$,$\theta_{obs}=\pi/2$]{
        \includegraphics[width=1.4in]{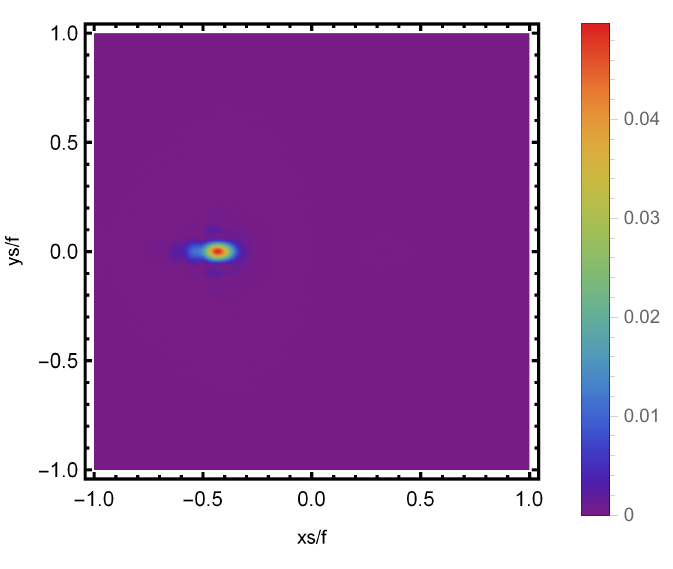}
    }
    \subfigure[$c_0=7$,$\theta_{obs}=0$]{
        \includegraphics[width=1.4in]{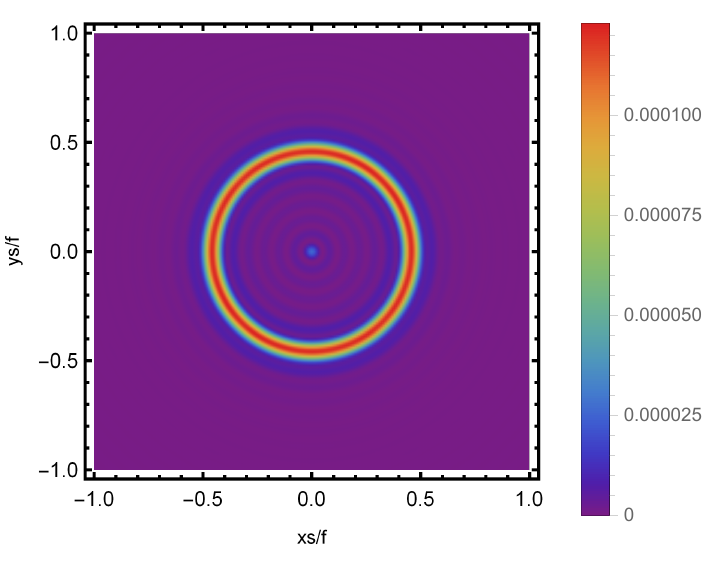}
    }
	\subfigure[$c_0=7$,$\theta_{obs}=\pi/6$]{
        \includegraphics[width=1.4in]{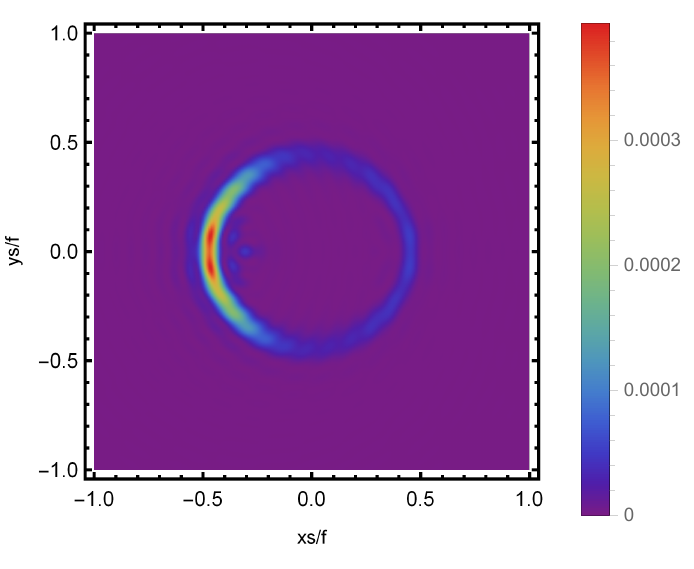}
    }
    \subfigure[$c_0=7$,$\theta_{obs}=\pi/3$]{
        \includegraphics[width=1.4in]{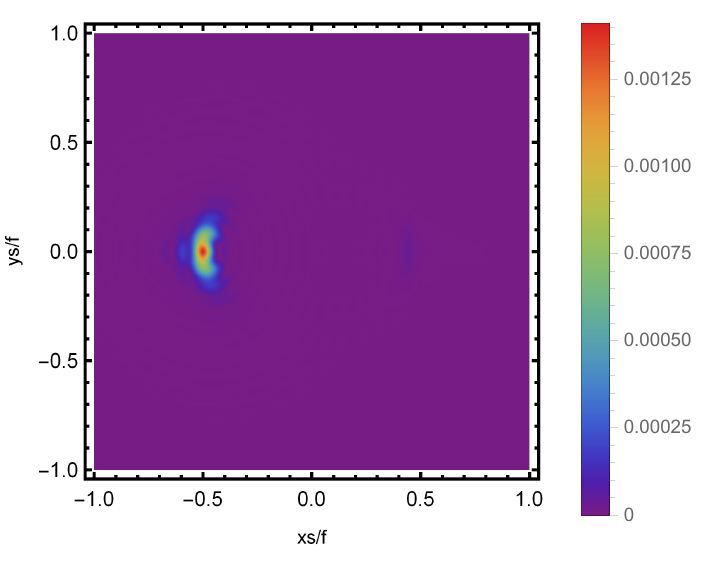}
    }
    \subfigure[$c_0=7$,$\theta_{obs}=\pi/2$]{
        \includegraphics[width=1.4in]{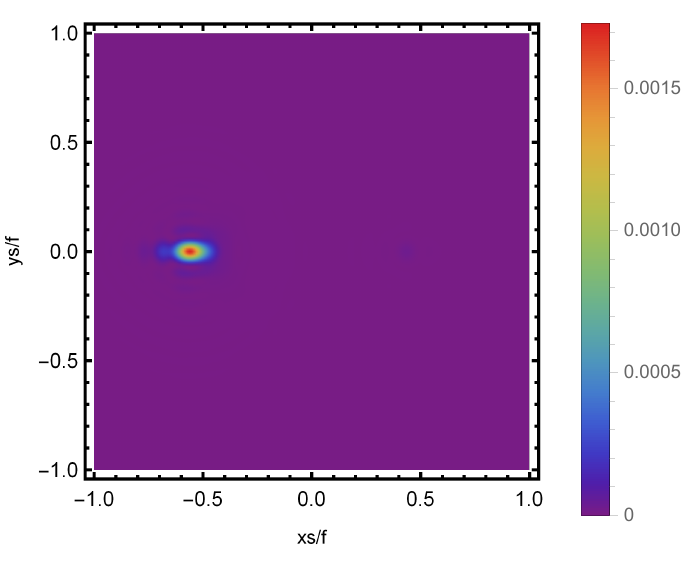}
    }
    \caption{The images of the lensed response are shown for different $c_0$  observed at various observation angles with $Q =4$, $z_h =4$,  $\omega = 80$, $e = 0.01$.}
    \label{image_complete}
\end{figure}

\begin{figure}
    \centering
    \subfigure[$c_0=1$]{
        \includegraphics[width=1.4in]{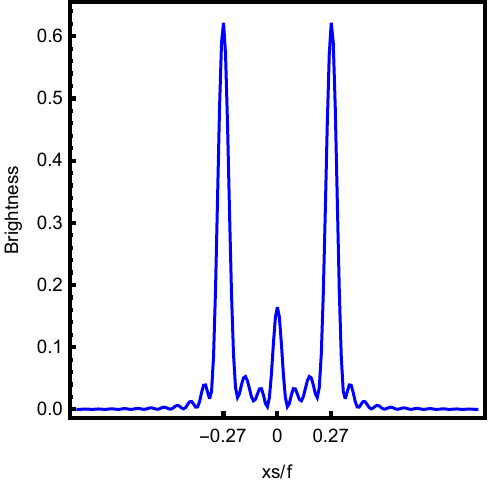}
    }
	\subfigure[$c_0=3$]{
        \includegraphics[width=1.4 in]{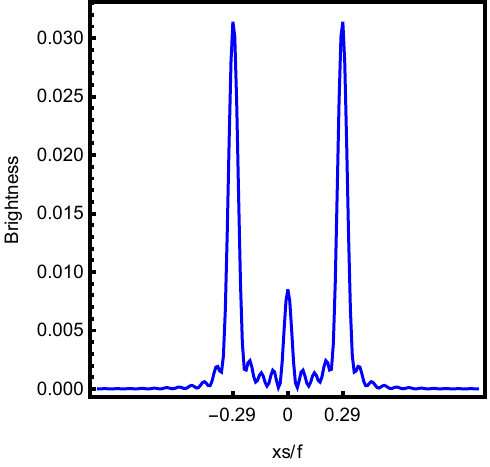}
    }
    \subfigure[$c_0=5$]{
        \includegraphics[width=1.4in]{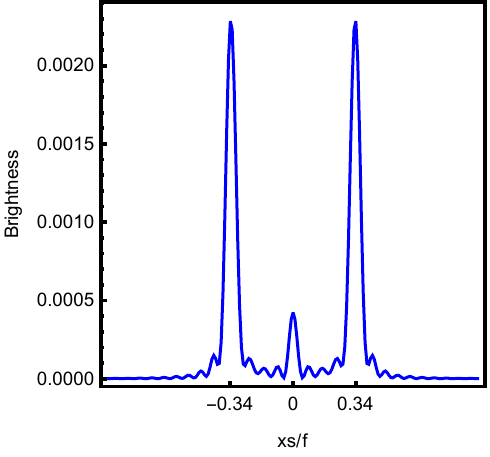}
    }
    \subfigure[$c_0=7$]{
        \includegraphics[width=1.4 in]{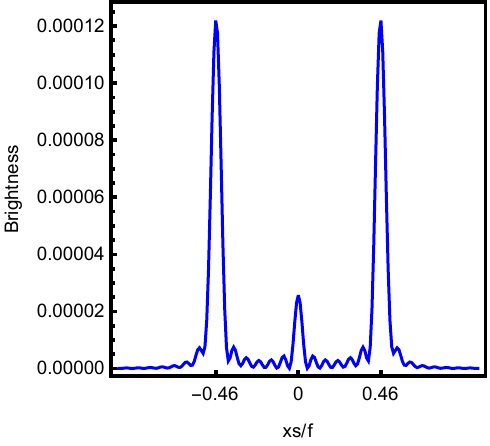}
    }
    \caption{The brightness of the lensed response for different $c_0$ observed at the observation angle $\theta_{obs}=0$  with $Q =4$, $z_h =4$, $\omega = 80$, $e = 0.01$.}
    \label{image_curve}
\end{figure}

Also, the images of the lensed response observed at various observation angles for different $c_0$ with $Q=4$, $r_{h*}=4$, $\omega=80$ and $e=0.01$ are plotted in Fig.~\ref{image_complete}. At first, the observer is at the position $\theta_{obs}=0^{\circ}$. This means the observation location is at the north pole of the AdS boundary. From the left-most column of Fig.~\ref{image_complete}, a series of axisymmetric concentric rings appear and they tend to move away from the center as $c_0$ increases. Next we fix our observer to $\theta=\pi/6$ (the second column from the left shown in Fig.~\ref{image_complete}). As the parameter $c_0$ increases, the same phenomena exists as in the $\theta=0^{\circ}$ case. Explicitly, there still exists a series of axisymmetric concentric rings. And the brightness on the right side of the ring is less than the brightness on the left side of the ring. When the observer moves to $\theta=\pi/3$, the luminosity rings disappear. All left are just bright light arcs or light spots which are consistent with~\cite{Zeng:2023zlf,Zeng_Li2023}. And these light spots or light arcs exist only on the left side. When the parameter $c_0$ increases, the bright light spots change to the light arcs. At last, we move to $\theta=\pi/2$. Only bright spots are shown on the right-most column of Fig.~\ref{image_complete}. As the parameter $c_0$ increases, the bright spot is far away from the center. 
To see this better, in Fig.~\ref{image_curve}, we plot the brightness for different 
 $c_0$  at $\theta_{obs}=0$. Obviously, the peak of the curves corresponds to the radius of rings. From these subfigures, we see that the ring radius increases correspondingly as $c_0$ increases.

\begin{figure}
    \centering
	\subfigure[$u=0.1$]{
        \includegraphics[width=1.4in]{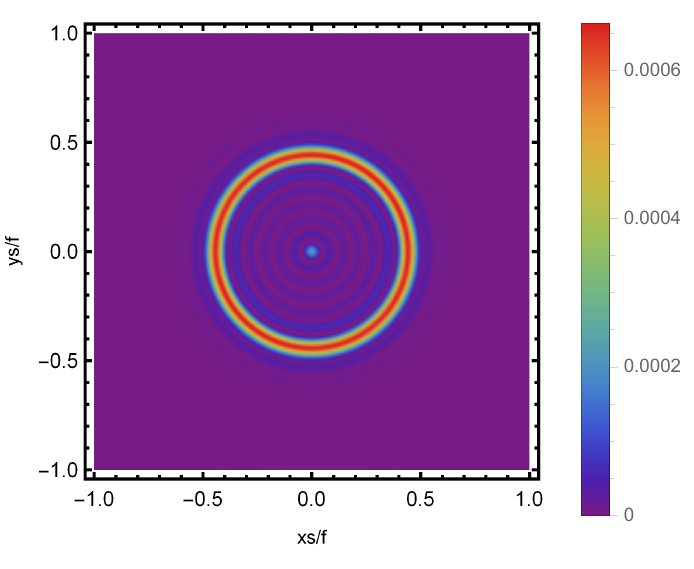}
    }
    \subfigure[$u=1.1$]{
        \includegraphics[width=1.4in]{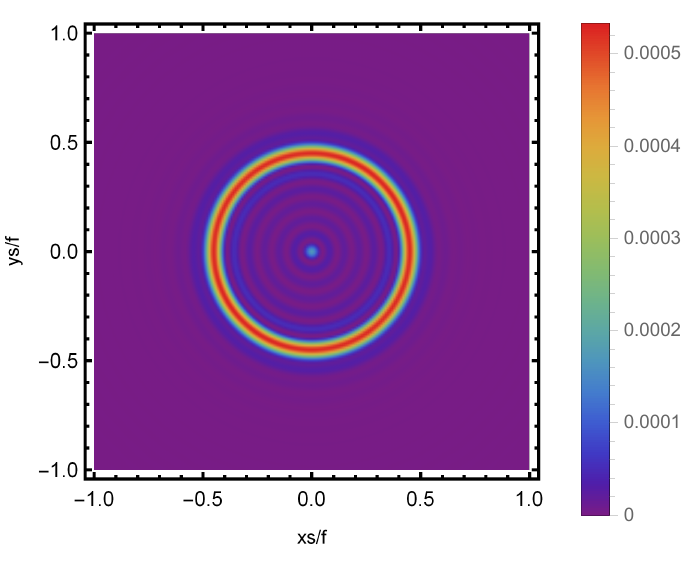}
    }
    \subfigure[$u=1.8$]{
        \includegraphics[width=1.4in]{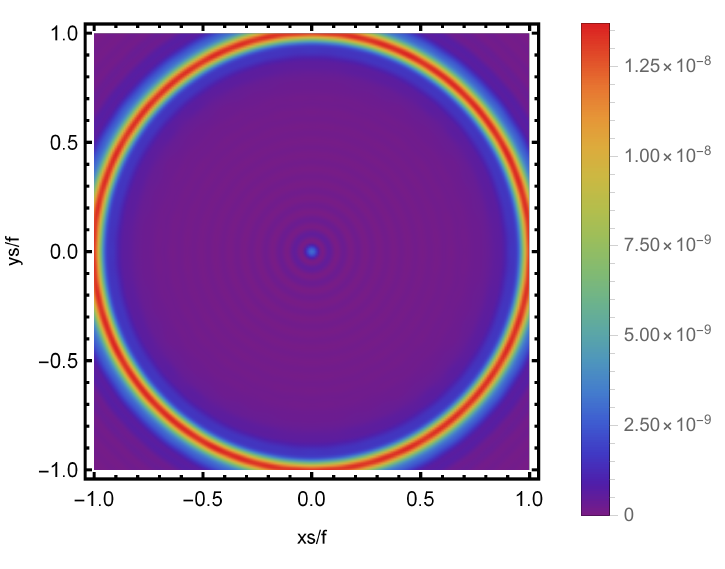}
    } \subfigure[$u=2.5$]{
        \includegraphics[width=1.4in]{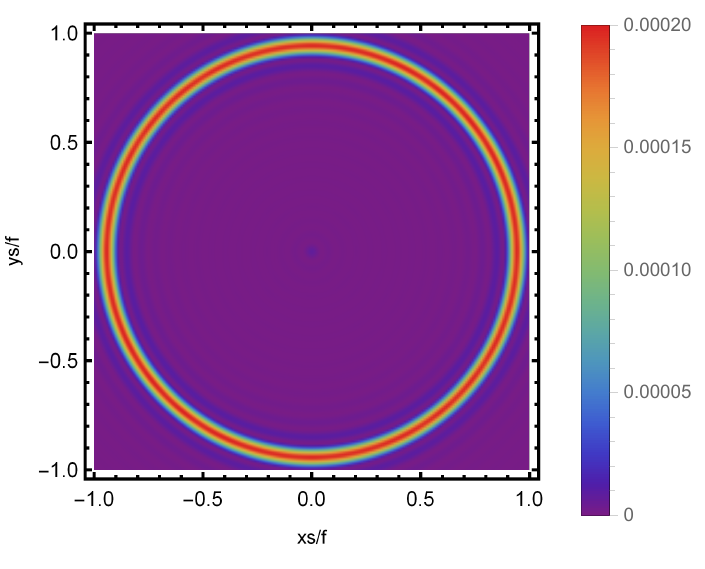}
    }
    \caption{The images of the lensed response for different chemical potential  $u$ observed at the observation angle $\theta_{obs}=0$ with a fixed temperature $T=0.5$. Here $\omega=80$ and  $c_0=0.5$, $e=0.01$.  }\label{sharp_3}
\end{figure}

\begin{figure}
    \centering
    \subfigure[$u=0.1$]{
        \includegraphics[width=1.4in]{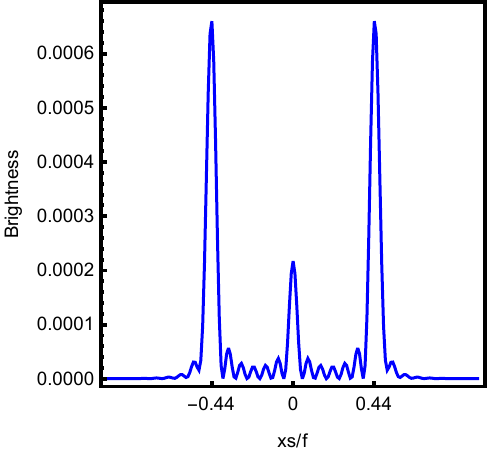}
    }
	\subfigure[$u=1.1$]{
        \includegraphics[width=1.4in]{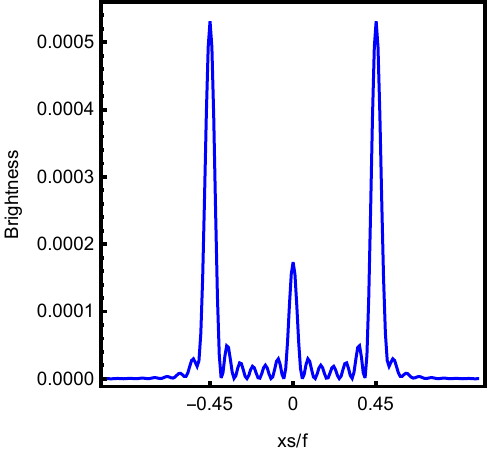}
    }
    \subfigure[$u=1.8$]{
        \includegraphics[width=1.4in]{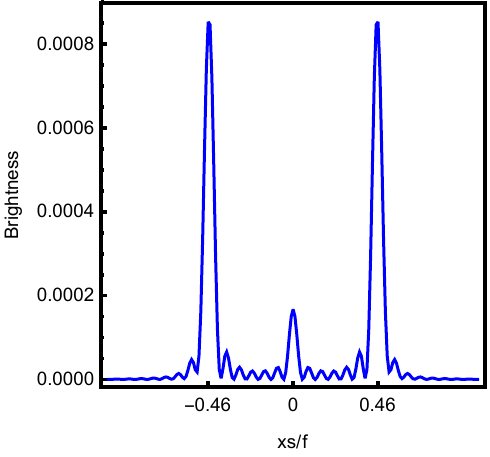}
    }
    \subfigure[$u=2.5$]{
        \includegraphics[width=1.4in]{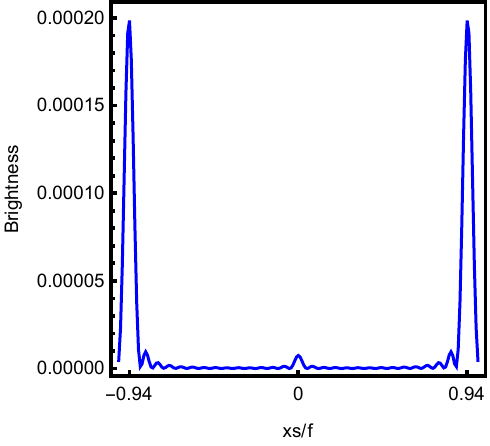}
    }
    \caption{The brightness of the lensed response for different chemical potential $u$ are shown on the screen with a fixed temperature $T=0.5$. Here $\omega=80$ and $c_0=0.5$, $e=0.01$.}
    \label{sharp_4}
\end{figure}

The images of the lensed response observed at the observation angle $\theta_{obs}=0$ for different chemical potentials $u$ with a fixed temperature $T=0.5$ are shown in Fig.~\ref{sharp_3}. Here we set $\omega=80$, $c_0=0.5$ and $e=0.01$. These figures clearly show that as the potential increases from $u=0.1$ to $u=1.8$, the positions of the brightest ring vary slightly. However, as the chemical potential $u$ increases to $u=2.5$, the brightest ring is further away from the center. These phenomena are also seen in Fig.~\ref{sharp_4} which shows the profile of the lensed response. The peaks of these curves change slowly from $u=0.1$ to $u=1.8$, but position of the brightness peak is far away from the center for  $u=2.5$. Nevertheless, from both Fig.~\ref{sharp_3} and  Fig.~\ref{sharp_4}, we know that as  $u$ increases, the ring radius increases, too.


\begin{figure}
    \centering
	 \subfigure[$T=0.914$]{
        \includegraphics[width=1.4in]{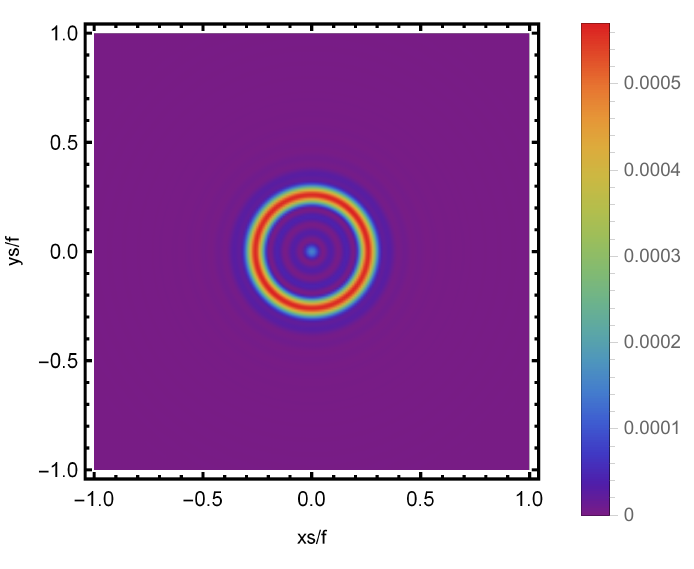}
    }
	\subfigure[$T=0.369$]{
        \includegraphics[width=1.4in]{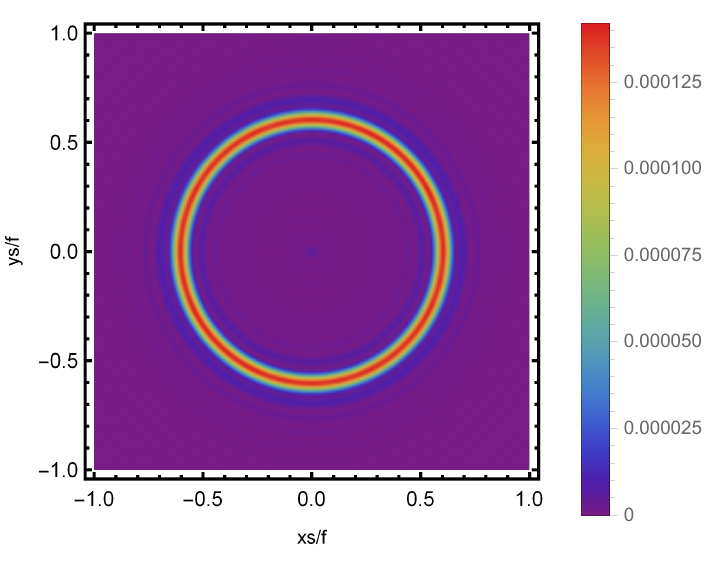}
    }
    \subfigure[$T=0.299$]{
        \includegraphics[width=1.4in]{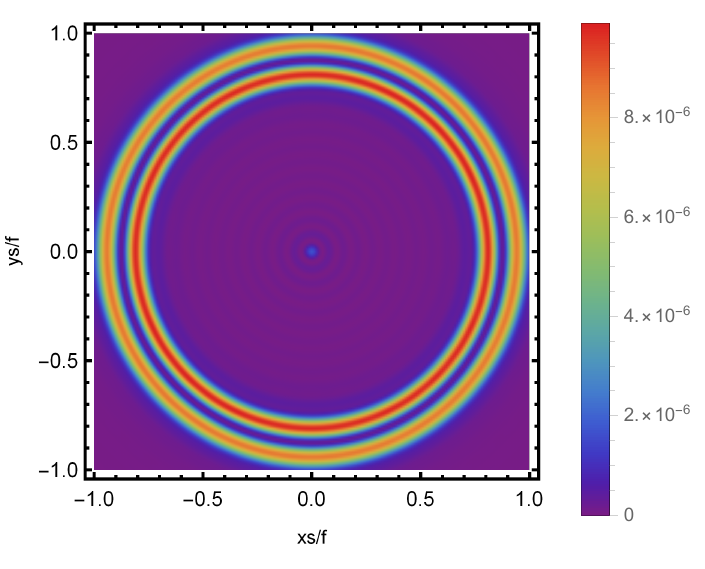}
    }
    \subfigure[$T=0.296$]{
        \includegraphics[width=1.4in]{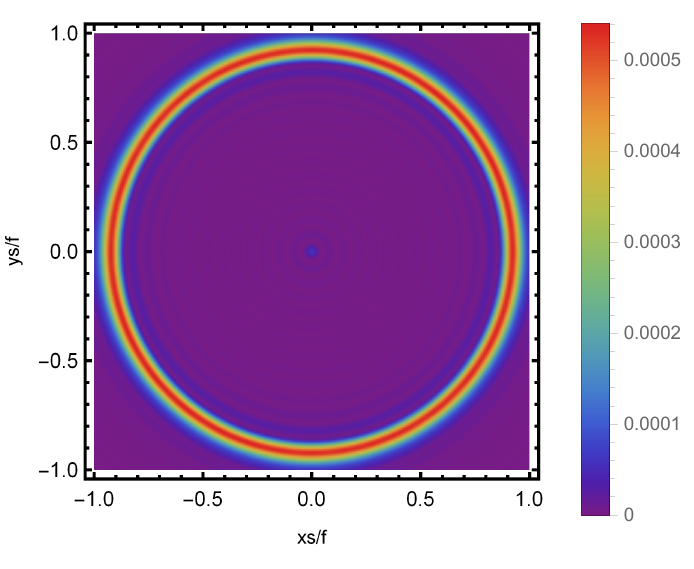}
    }
    \caption{The images of the lensed response for different $T$ observed at the observation angle $\theta_{obs}=0$ with a fixed $u=1$. Here, $\omega=80$, $c_0=0.5$, $e=0.01$ and from (a) to (d), the charges correspond to $Q=0.1, 0.3, 0.5, 0.7$ respectively.}
    \label{temperature_1}
\end{figure}

\begin{figure}
    \centering
    \subfigure[$T=0.914$]{
        \includegraphics[width=1.4in]{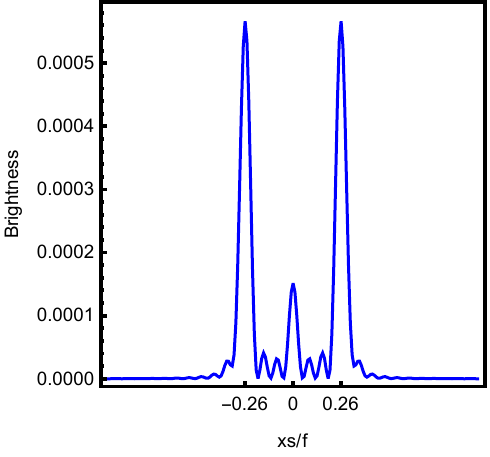}
    }
	\subfigure[$T=0.369$]{
        \includegraphics[width=1.4in]{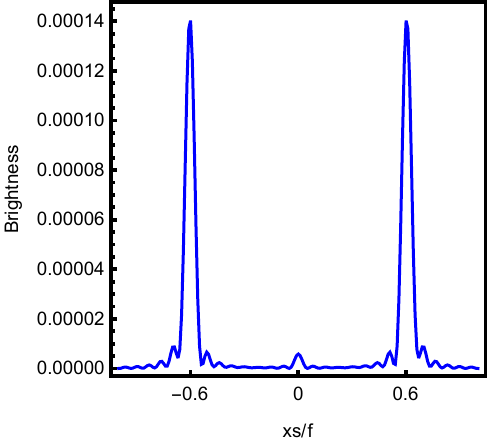}
    }
    \subfigure[$T=0.299$]{
        \includegraphics[width=1.4in]{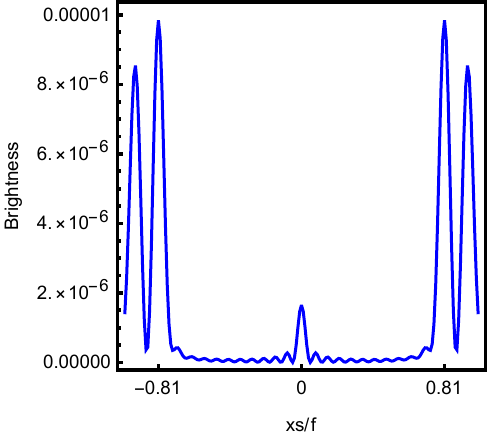}
    }
    \subfigure[$T=0.296$]{
        \includegraphics[width=1.4in]{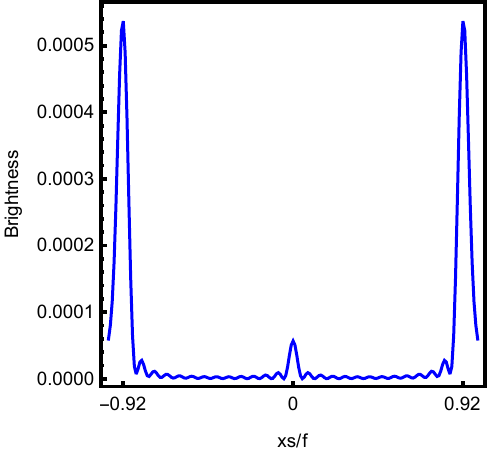}
    }
    \caption{ The brightness of the lensed response are shown on the screen for different $T$ with a fixed $u=1$. Here, $\omega=80$, $c_0=0.5$, $e=0.01$ and from (a) to (d), the charges correspond  to $Q=0.1, 0.3, 0.5, 0.7$ respectively.}
     \label{temperature_2}
\end{figure}

We also investigate the effect of the horizon temperature $T$ on the profiles of the lensed response shown in Fig.~\ref{temperature_1}, which is observed at $\theta_{obs}=0$ with fixed parameters $u=1$, $\omega=80$, $c_0=0.5$ and $e=0.01$. We depicted the observational image of the dual  black hole  for increasing values of temperature $T$. When the charges $Q=0.1,0.3,0.5,0.7$, the corresponding temperatures $T=0.914, 0.369, 0.299, 0.296$ respectively. We see one particular bright ring in the region. When the temperature decreases from $T=0.914$ to $T=0.296$, the brightest ring moves far away from the center, which is also seen from the curves in Fig.~\ref{temperature_2}, where the peak is also far away from the center as the temperature $T$ decreases.

\section{The comparison between the holographic results and optical results}
The above analysis shows that the brightest ring exists at the position of the photon sphere of the black holes. We will verify this bright ring existing here from the optical geometry. 

 
 It is well-known that in the Eikonal approximation, the Klein-Gordon equation reduced to the following  Hamilton-Jacoboi equation  
 \begin{equation}
	g^{\mu \nu} (\partial _\mu S-e A_t)  (\partial _\nu S-e A_t)-M^2=0, 
\end{equation}
 in which $S$ is the action. 
For a spacetime with metric in Eq.(\ref{metric_metric}), 
This equation can be expressed as
\begin{equation}
	-\frac{1}{F(r)}\left(\partial _t S-e A_t \right )^2 +F(r)\partial _r S^2+\frac{1}{r^2}\partial _{\theta}^2+\frac{1}{r^2 \sin^2 \theta}\partial _{\phi}^2-M^2=0.
\end{equation}
Next we only consider  $\theta=\pi/2$,  so that the photon orbit lying on the equatorial plane. For the static spherically symmetric
 space time, there are two killing vectors, so that the action can be written as 
  \begin{equation}
S(t,r, \phi)=-\omega t+ L \phi+\int \frac{1}{F(r)}\sqrt{R} dr,
\end{equation}
in which 
 \begin{equation}
 R= (\omega-e A_t)^2-F(r)\left (\frac{L^2}{r^2}-2\right),
\end{equation}
and 
 \begin{equation}
\frac{\partial S}{\partial t}=-\omega,\frac{\partial S}{\partial \phi}=L, \frac{\partial S}{\partial r}=\frac{\sqrt{R}}{F(r)}.
\end{equation}
\begin{figure}[ht]
	\centering
	\includegraphics[trim=1.2cm 0.5cm 4.9cm 2.5cm, clip=true, scale=0.9]{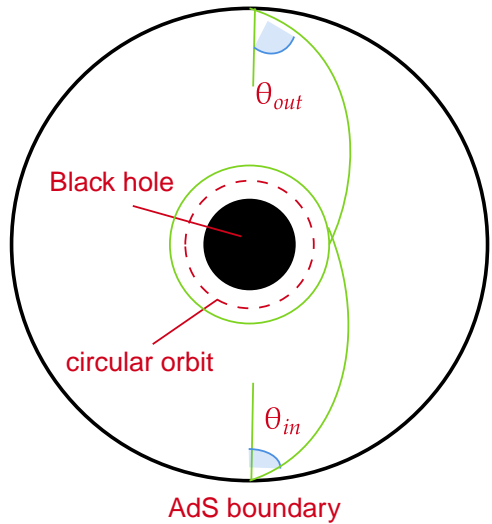}
	\caption{The ingoing angle and outgoing angle of the photon at the photon sphere.}
 \label{ingoing_and_outgoing_1}
\end{figure}
The photon will propagate along the trajectory of geodesics satisfying the above conditions.

According to the definition ~\cite{
Hashimoto:2018okj,Hashimoto:2019jmw}
\begin{eqnarray}
\text{}\cos\theta_{in}&=&\frac{g_{ij}u^i n^j}{|u||n|}|_{r=\infty},
\end{eqnarray}
the ingoing angle $\theta_{in}$ with the normal vector of boundary $\text{}n^b=\frac{\partial }{\partial r^b}$ and four-velocity  $\text{}u^a=(\frac{\partial }{\partial \lambda})^a$  satisfies  
\begin{eqnarray}
\text{}\sin\theta_{in}=\frac{{L}}{\hat{\omega}}\text{},
\end{eqnarray}
which is shown in Fig.~\ref{ingoing_and_outgoing_1}. The above relation is still valid when the  photon is located at the photon ring. Label the angular momentum as $L_p$, which is determined by the following conditions ~\cite{Liu:2022cev}
\begin{equation}
R=0, \frac{d {R}}{dr}=0.
\end{equation}
In the geometrical optics, the angle $\theta_{in}$ gives the angular distance of the image of the incident ray from the zenith if an observer located on the AdS boundary looking up into the AdS bulk. When two end points of the geodesic and the center of the black hole are in alignment, the observer watches a ring with a radius which corresponds to the incident angle $\theta_{in}$ because of axial symmetry~\cite{Hashimoto:2018okj,Hashimoto:2019jmw}. 
\begin{figure}[ht]
	\centering
	\includegraphics[trim=1.2cm 1.5cm 0.4cm 0.1cm, clip=true, scale=0.9]{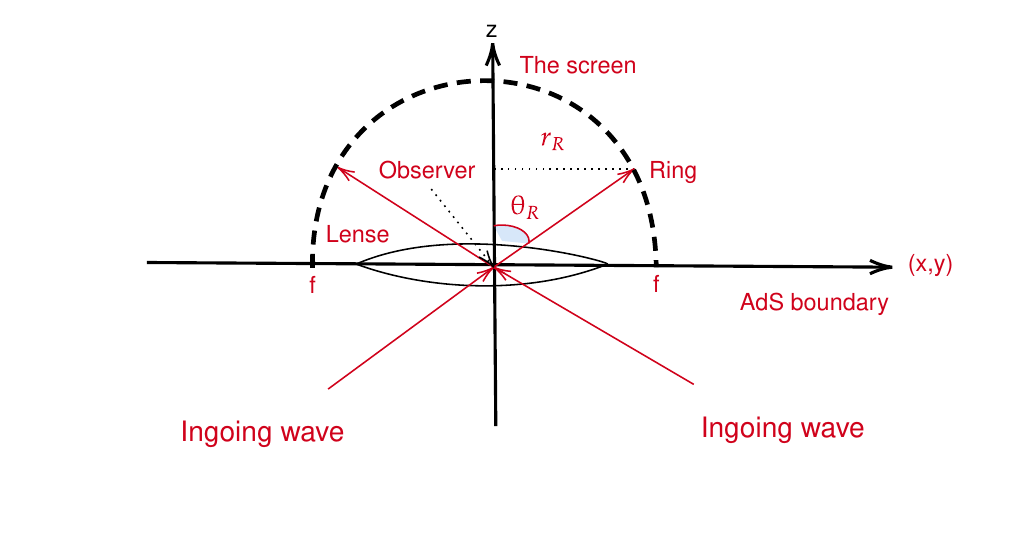}
	\caption{The  diagram   between the ring angle  $\theta_R $ and ring radius  $r_R$.}\label{irreg11111}
\end{figure}

\begin{figure}
    \centering
    \subfigure[$c_0=-0.5$]{
        \includegraphics[width=1.9in]{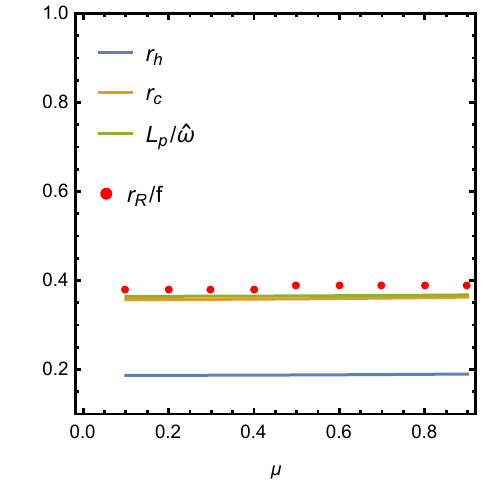}
    }
	\subfigure[$c_0=0.5$]{
        \includegraphics[width=1.9in]{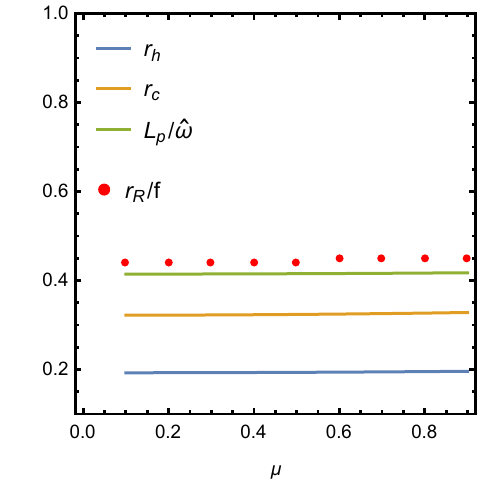}
    }
    \subfigure[$c_0=1.5$]{
        \includegraphics[width=1.9in]{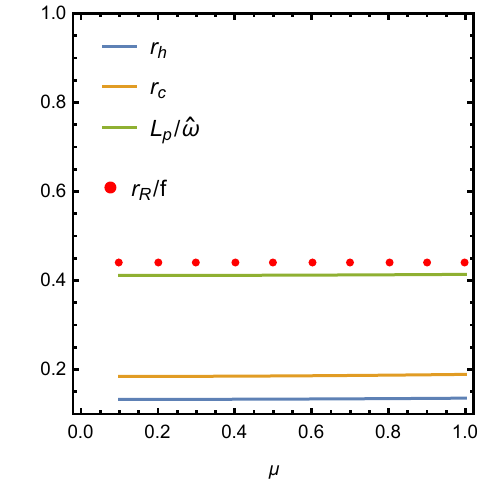}
    }
    \caption{Comparison between  the results obtained by wave optics and geometric optics for different $c_0$ with a fixed temperature $T=0.5$.}
    \label{comparison_1}
\end{figure}

\begin{figure}
    \centering
    \subfigure[$c_0=-0.5$]{
        \includegraphics[width=1.9in]{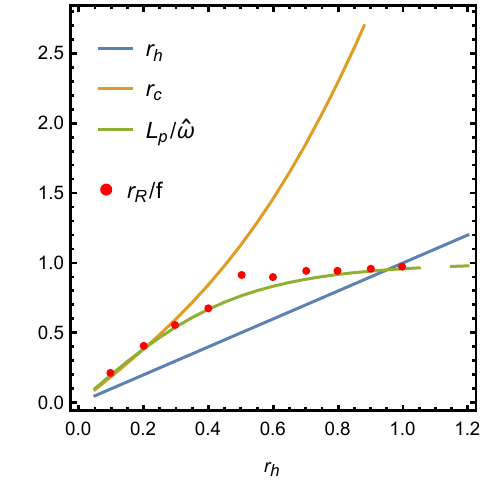}
    }
	\subfigure[$c_0=0.5$]{
        \includegraphics[width=1.9in]{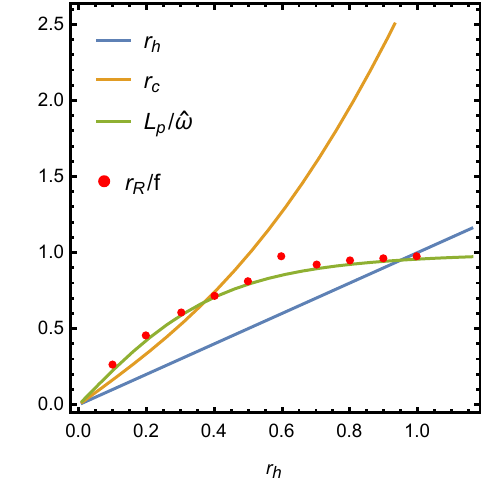}
    }
    \subfigure[$c_0=1.5$]{
        \includegraphics[width=1.9in]{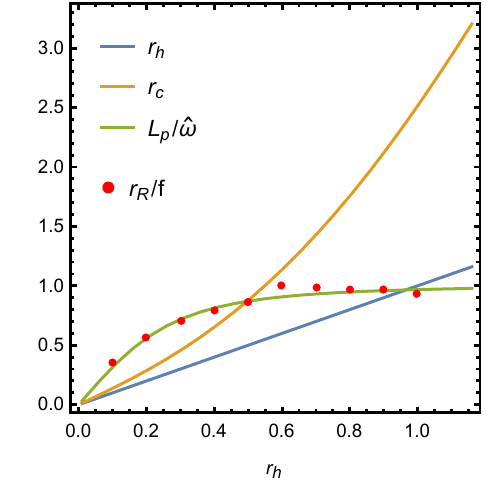}
    }
    \caption{Comparison between  the results obtained by wave optics and geometric optics for different $c_0$  with a fixed chemical potential $u$.}
     \label{comparison_2}
\end{figure}

Furthermore, with Fig.~\ref{comparison_2}, the angle of the Einstein ring is 
\begin{equation}
\text{}\sin \theta_R=\frac{r_R}{f}\text{}.
\end{equation}\text{}
Following~\cite{Hashimoto:2018okj,Hashimoto:2019jmw}, we have $\sin\theta_R=\sin \theta_{in}$ for a sufficiently large $l$. Then we obtain 
\begin{equation}
\text{}\frac{r_R}{f}=\frac{L_p}{\hat{\omega}},\text{}
\end{equation}
where $L_p$ is the angular momentum at the photon sphere. This relation also can be further proved numerically. In Fig. \ref{comparison_1} and Fig.~\ref{comparison_2}, we fit $\frac{r_R}{f}$ with $\frac{L_p}{\hat{\omega}}$ for different chemical potential and horizon respectively. Both
radii of the black hole horizon and the circular orbit are also exhibited simply for curiosity’s sake. From both figures, we can see that 
the Einstein ring angle derived by our holographic method is consistent well with the ingoing angle obtained by geometric optics.

\begin{figure}
    \centering
    \subfigure[]{
        \includegraphics[width=2.5in]{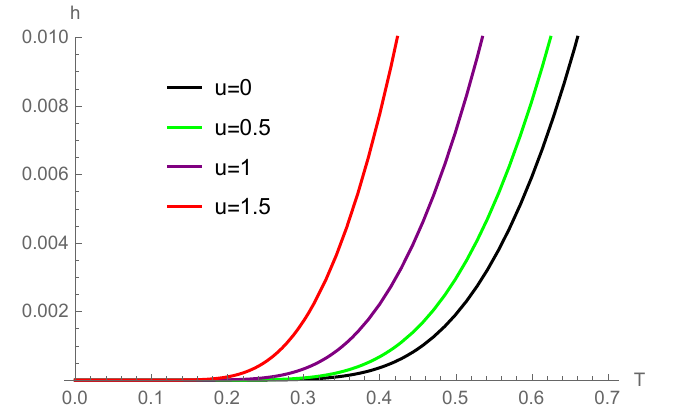}
    }
	\subfigure[]{
        \includegraphics[width=2.5in]{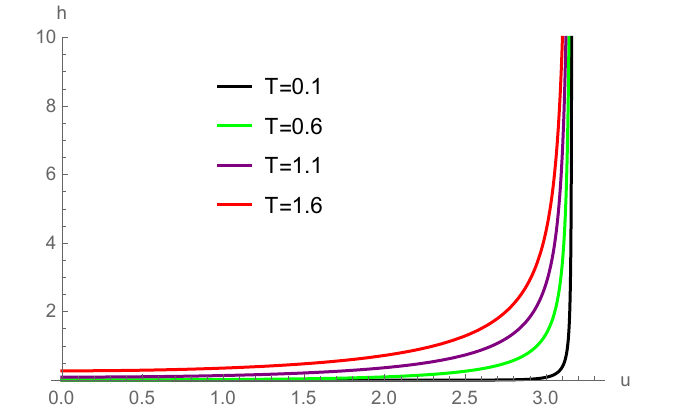}
    }
    \caption{ The dependence of the correction term $h$ on the temperature and the chemical potential
for $m^2=10$.}   \label{vh}
\label{last_figre}
\end{figure}

In fact, the  response function or the Einstein ring  can be calculated directly by the Green function
\begin{equation} \text{}\text{}\text{}
    \langle\mathcal{O}\rangle_{J_\mathcal{O}}=\int_0^{2\pi} \text{}\text{}\text{}\varphi'\int_0^\pi \theta'\sin\theta' G(t,\theta,\varphi;t',\theta',\varphi')J_\mathcal{O}\text{}\text{}\text{}(t',\theta'\varphi')=\sum_{l=0}^\infty e^{-i\omega t}G_{l0}(\omega)c_{l0}Y_{l0}(\theta).\text{}\text{}\text{}
\end{equation}
For a weakly interacting quantum field theory at finite temperature and finite chemical potential, the retarded Green function can be expressed as  ~\cite{Liu:2022cev}
\begin{equation}\text{}
       \text{} G_{lm}(\omega)=\frac{1}{\hat{\omega}^2-m_T^2-l(l+1)},\text{}
\end{equation}
where the thermal mass is given by $m_T^2=m^2+\frac{\gamma}{2} h(T,u)$, in which 
\begin{equation}
    h=\sum_l\frac{1}{\omega_l}[\frac{1}{e^{\beta(\omega_o+u)}-1}+\frac{1}{e^{\beta(\omega_o-u)}-1}],
\end{equation}
where $\beta$ is the inverse of the temperature and $\omega_o=\sqrt{m^2+l (l+1)}$. Principally, as we know the variation tendency of $h$, we can know the response or ring radius qualitatively.  In Fig.~\ref{vh}, we plot  $h$ with respect to $T$ and $u$. 
From (a), we know that for the  low temperature, $h$ changes little. While for high temperature, we  find the higher the   $u$, the larger the $h$. In other words, the response or ring radius increases as the potential $u$ increases. This result is consistent with that obtained by  Fig.~\ref{sharp_4}. From (b), we know that for the small $u$, the higher the $T$ , the larger the  $h$. That is to say, as the temperature $T$ raises, the  ring radius increases too. This conclusion is inconsistent with that obtained by Fig.~\ref{temperature_2}, but consistent with reference~\cite{Liu:2022cev}.

\section{Conclusions}
\label{sec5}
As well known, the black holes are one of the incredible predictions of General Relativity. And this gives us an opportunity to explore the formation of black hole shadows. The Event Horizon Telescope collaboration obtained an amazing success by releasing the first-ever image of a black hole. The Event Horizon Telescope is important for deeper understanding and explanation of the observed phenomena, such  as Einstein ring. Therefore, in this paper, with the holographic dual, we study the Einstein ring of the charged black hole in conformal gravity. 

Usually, the real quantum materials are engineered   at a finite
chemical potential.  To  realize this in the framework of holography, we should turn on the bulk electromagnetic field. Therefore, the black hole should be charged. We investigated firstly the effect of the chemical potential on the  amplitude of $\langle O \rangle$.   We found that when the chemical potential increases, the amplitude of $\langle O \rangle$ increases correspondingly. In addition, we also investigated the effect of the chemical potential on the ring radius, and it was found that the ring radius increases as the chemical potential increases though it changes a little for the low chemical potential. 
Moreover, the effect of the temperature on the  amplitude of $\langle O \rangle$  and ring radius are discussed too. We found the higher the temperature, the smaller the  amplitude of $\langle O \rangle$, and the higher the temperature, the smaller the ring radius. Though the amplitude of $\langle O \rangle$ does not describe the Einstein ring, the change tendency is the same as that of the ring radius. 

Conformal gravity is an important   theories of gravity partly. On one hand,  its
on-shell equivalence to Einstein gravity and its power counting renormalizability. On the other hand, 
 its spherically symmetric solution contains a term
linear in radial coordinate, which may  explain the galaxy rotation curves. We hence also investigated the effect of the  gravity-related parameter $c_0$  on the  amplitude of $\langle O \rangle$ as well as  the  ring radius, and found they grows as  $c_0$  increases. 

All the results are obtained in the framework of holography. To verify the correctness of these results, we also obtained the   radius of photon ring  and the corresponding ingoing angle via geometric optics. For different temperatures and chemical potentials, we  found the ingoing angles are consistent with the angles of Einstein ring.   
The wave optics in the framework of holography thus is  reliable. 

In \cite{Liu:2022cev}, Einstein rings of a charged black hole in Einstein gravity were investigated.  They found  the chemical potential has little effect on the ring radius. They considered maybe only the small chemical potential. As the chemical potential is large enough, its effect on the ring radius will be obvious. In addition, in~\cite{Liu:2022cev}, 
it was found that the higher the temperature, the smaller the ring radius. Obviously, our result is opposite to this result.  The reason for this is that in conformal gravity, there is no charge in the metric, but it is given by the constraint relation. Therefore, the holographic
images play an important role in differentiating the geometric features of different black holes for
fixed wave source and optical system.

\section*{Acknowledgements}{This work is supported  by the National
Natural Science Foundation of China (Grants No. 11675140, No. 11705005, and No. 12375043), and  Innovation and Development Joint  Foundation of Chongqing Natural Science  Foundation (Grant No. CSTB2022NSCQ-LZX0021) }and Basic Research Project of Science and Technology Committee of Chongqing (Grant No. CSTB2023NSCQ-MSX0324).


\begin{thebibliography}{99}

\bibitem{Giesler2019}
M. Giesler, M. Isi, M. A. Scheel, and S. Teukolsky, Black Hole Ringdown: The Importance of Overtones, Phys. Rev. X
9, 041060 (2019)

\bibitem{Isi2019}
M. Isi, M. Giesler, W. M. Farr, M. A. Scheel, and S. A. Teukolsky, Testing the no-hair theorem with GW150914, Phys.
Rev. Lett. 123, 111102 (2019)

\bibitem{Isi2021} 
M. Isi, W. M. Farr, M. Giesler, M. A. Scheel, and S. A.
Teukolsky, Testing the Black-Hole Area Law with GW150914,
 Phys. Rev. Lett. 127, 011103 (2021)

\bibitem{Isi_2021} 
M.~Isi and W.~M.~Farr,
Analyzing black-hole ringdowns,
[arXiv:2107.05609 [gr-qc]]

\bibitem{Franchini2023} 
N. Franchini and S. H. Volkel, Testing General Relativity with Black Hole Quasi-Normal Modes,   arXiv:2305.01696
 [gr-qc]

\bibitem{Nicolini09}
P. Nicolini, Noncommutative Black Holes, The Final Appeal To Quantum Gravity: A Review,
Int. J. Mod. Phys. A 24, 1229-1308 (2009)

\bibitem{Bamber2021}
 J. Bamber, O. J. Tattersall, K. Clough, and P. G. Ferreira, Quasinormal modes of growing dirty black holes, Phys. Rev.
 D 103, 124013 (2021)

\bibitem{Cardoso2022}
 V. Cardoso, K. Destounis, F. Duque, R. P. Macedo, and A. Maselli, Black holes in galaxies: Environmental impact on
 gravitational-wave generation and propagation, Phys. Rev. D 105, L061501 (2022)

\bibitem{Barack2019}
 L. Barack et al., Black holes, gravitational waves and fundamental physics: a roadmap, Class. Quant. Grav. 36, 143001
 (2019)

\bibitem{Meszaros2019_astro}
 P. Meszaros, D. B. Fox, C. Hanna, and K. Murase, Multi-Messenger Astrophysics, Nature Rev. Phys. 1, 585 (2019)

 
\bibitem{Ashtekar2014}
A. Ashtekar, M. Reuter, and C. Rovelli, From General Relativity to Quantum Gravity, arXiv:1408.4336 [gr-qc]

\bibitem{Pawlowski2021}
J. M. Pawlowski and M. Reichert, Quantum Gravity: A Fluctuating Point of View, Front. in Phys. 8, 551848
 (2021)

 \bibitem{Akiyama2019}
K. Akiyama et al., [Event Horizon Telescope Collaboration] First M87 event horizon telescope results. I. The shadow of
 the supermassive black hole, Astrophys. J. Lett. 875, L1 (2019) 

\bibitem{Akiyama2019Lett} 
K. Akiyama et al., [Event Horizon Telescope Collaboration], First M87 event horizon telescope results. II. Array and
 instrumentation, Astrophys. J. Lett. 875, L2 (2019) 
 
\bibitem{Akiyama2019Astrophys}  
K. Akiyama et al., [Event Horizon Telescope Collaboration], First M87 event horizon telescope results. III. Data processing
 and calibration, Astrophys. J. Lett. 875, L3 (2019) 

\bibitem{Akiyama2019_Lett}   
K. Akiyama et al., [Event Horizon Telescope Collaboration], First M87 event horizon telescope results. IV. Imaging the
 central supermassive black hole, Astrophys. J. Lett. 875, L4 (2019) 
 
\bibitem{Akiyama2019_ring}    
K. Akiyama et al., [Event Horizon Telescope Collaboration], First M87 event horizon telescope results. V. Physical origin
 of the asymmetric ring, Astrophys. J. Lett. 875, L5 (2019) 

\bibitem{Akiyama2019_875}   
K. Akiyama et al., [Event Horizon Telescope Collaboration], First M87 event horizon telescope results. VI. The shadow
 and mass of the central black hole, Astrophys. J. Lett. 875, L6 (2019) 

\bibitem{Gralla2019}  
S. E. Gralla, D. E. Holz and R. M. Wald, Black Hole shadows, photon rings, and lensing rings, Phys. Rev. D 100, 024018
 (2019) 

\bibitem{Zeng2023}  
X. X. Zeng, M. I. Aslam and R. Saleem, The Optical Appearance of Charged Four-Dimensional Gauss-Bonnet Black Hole
 with Strings Cloud and Non-Commutative Geometry Surrounded by Various Accretions Profiles, Eur. Phys. J. C 83, 129
 (2023) 
 
\bibitem{Zeng2020} 
X. X. Zeng and H. Q. Zhang, Influence of quintessence dark energy on the shadow of black hole, Eur. Phys. J. C 80, 1058
 (2020) 

\bibitem{Zeng:2020vsj}
X.~X.~Zeng and H.~Q.~Zhang,
Influence of quintessence dark energy on the shadow of black hole,
Eur. Phys. J. C 80, 1058  (2020)  

\bibitem{Hashimoto:2018okj}
K.~Hashimoto, S.~Kinoshita and K.~Murata,
maging black holes through the AdS/CFT correspondence,
Phys. Rev. D  101, 066018 (2020)

\bibitem{Hashimoto:2019jmw}
K.~Hashimoto, S.~Kinoshita and K.~Murata,
Einstein Rings in Holography
Phys. Rev. Lett. 123, 031602 (2019)

\bibitem{Liu:2022cev}
Y.~Liu, Q.~Chen, X.~X.~Zeng, H.~Zhang, W.~L.~Zhang and W.~Zhang,
Holographic Einstein ring of a charged AdS black hole,
JHEP 10, 189 (2022)

\bibitem{Zeng:2023zlf}
X.~X.~Zeng, K.~J.~He, J.~Pu and G.~P.~Li,
Holographic Einstein rings of a Gauss-Bonnet AdS black hole,
[arXiv:2302.03692 [gr-qc]]

\bibitem{Zeng_Li2023}
X.~X.~Zeng, L.~F.~Li and P.~Xu,
Holographic Einstein rings of a black hole with a global monopole,
[arXiv:2307.01973 [hep-th]]


\bibitem{Hu:2023eow}
X.~Y.~Hu, X.~X.~Zeng, L.~F.~Li and P.~Xu,
Holographic Einstein rings of Non-commutative black holes,
[arXiv:2309.07404 [gr-qc]].


\bibitem{CEJM}
A. Chamblin, R. Emparan, C. V. Johnson, and R. C. Myers, Phys. Rev.  D 60 , 064018 (1999) 


\bibitem{1}   P. D. Mannheim, Making the case for conformal gravity. Found. Phys 42, 388  
 (2012)

 
\bibitem{2} E.~Bergshoeff, M.~de Roo and B.~de Wit,
Extended Conformal Supergravity,
Nucl. Phys. B  182  173-204  (1981)


\bibitem{3} P.~D.~Mannheim and J.~G.~O'Brien,
Galactic rotation curves in conformal gravity,
J. Phys. Conf. Ser.  437  012002 (2013)


\bibitem{4} R. K. Nesbet, Conformal Gravity: Dark Matter and Dark Energy, Entropy 15, 162-176 (2013)



\bibitem{Li:2012gh}
J.~Li, H.~S.~Liu, H.~Lu and Z.~L.~Wang,
Fermi Surfaces and Analytic Green's Functions from Conformal Gravity,
JHEP  02, 109 (2013)

\bibitem{Falcke:1999pj}
  H.~Falcke, F.~Melia and E.~Agol,
  Viewing the shadow of the black hole at the galactic center,
  Astrophys.\ J.\    528, L13  (2000)

\bibitem{Sharma2006}
K. Sharma, Optics: principles and application, Academic Press, 2006 
\end{thebibliography}

\end{document}